# Review of Flicker Noise Spectroscopy in Electrochemistry

Serge F. Timashev[a], Yuriy S. Polyakov[b*]

[a]*Karpov Institute of Physical Chemistry, Moscow 103064, Russia*
[b]*USPolyResearch, Ashland, PA 17921, USA*

**Abstract**

Extraction of information from real signals with chaotically varying components is often a necessary step in the analysis of various physicochemical, electrochemical, and surface phenomena and structures. The fundamental problems of information extraction from chaotic data are how to separate actual information from the interfering noise and what practical calculation procedures are most adequate for this purpose. This review presents the fundamentals of Flicker-Noise Spectroscopy (FNS), a general phenomenological methodology in which the dynamics and structure of complex nonlinear systems are analyzed by extracting information from various chaotic signals generated by the systems. The basic idea of FNS is to treat the correlation links present in sequences of different irregularities, such as spikes, "jumps", and discontinuities in derivatives of different orders, on all levels of the spatiotemporal hierarchy of the system under study as main information carriers. The tools to extract and analyze the information are power spectra and difference moments (structural functions) of various orders. The expressions for these functions derived for the stationary case are used to determine the information parameters that characterize the loss of "memory" (correlation links) for various irregularities. The expressions derived for the nonstationary case are used to evaluate the indicators (factors) of nonstationarity that allow one to determine the time moments when the most noticeable changes in the state occur, especially the ones that precede catastrophic events. Presently, FNS can be applied to three types of problems: (1) determination of parameters or patterns that characterize the dynamics or structural features of open complex systems; (2) finding precursors of abrupt changes in the state of various open dissipative systems based on a priori information about the dynamics of the systems; and (3) determination of flow dynamics in distributed systems based on the analysis of dynamic correlations in chaotic signals that are simultaneously measured at different points in space. Examples of FNS applications to such problems as parameterization of AFM images for the surfaces of LiF single crystals and AFM images for surfaces with dendritic structures; determination of precursors for the electric breakdown in thin porous silicon films, determination of precursors for major earthquakes; and study of membrane-potential "interelectrode" correlations in electromembrane systems with overlimiting current densities, as well as to some other problems in electrochemistry, surface science, and medicine are discussed.

[*] Corresponding author. Tel.: +1-570-875-3353.
*E-mail address*: ypolyakov@uspolyresearch.com (Yu.S. Polyakov).

**Contents**

# 1. Introduction

Chaotic time and space series of dynamic variables that arise in studies of various natural processes and structures are often an important source of information about the system state and features of its evolution and structure [1]. For example, the kinetics of many physicochemical, electrochemical, corrosion, and bioelectrochemical processes is accompanied by noise generation, which virtually represents the dynamic fluctuations of electrical potential or current [2–7]. Though the problems of the origin and information content of electrochemical noise are widely discussed [2], the noise is often considered as an artifact, and the experimental data are smoothed out to remove the chaotic components attributed to such noise. It is evident that some of the fluctuations in electrical parameters are in fact related to the "noise" introduced by the measuring equipment. At the same time, many studies, starting with the pioneering work of Tyagay, demonstrate that electrochemical processes at electrode-electrolyte interfaces are one of the main sources of fluctuations in the values of electrode potential or electrical current [8–11]. This is attributed to the complexity of even the simplest elementary processes involving electron transfer, such as hydrogen reduction on platinum [12]. It is well known that the development of pitting corrosion of metals and semiconductors is accompanied by the generation of noise leading to dynamic irregularities in the variations of anode potential and electrical current density [4].

To learn about the state of a natural system by analyzing its chaotic series, one needs to know what information is hidden in the noise and how to extract it in practice. This immediately brings about additional questions. Can the standard noise smoothing procedures be applied to the analysis of chaotic signals? Is it possible at all to extract information from the noise in view of the fact that any complex system, especially natural, is characterized by virtually infinite dimensionality; that is, there are an infinite number of dynamic variables that determine the state of such system? Should we restrict our attention to obtaining only the averaged characteristics of the system under study using the postulates of statistical physics and consider this system as an element of some ensemble? But this would contradict one of the inherent properties of natural systems, their individuality and lack of repeatability in their evolution.

Luckily, the modern understanding of complex systems dynamics and related computer simulations tell us that the presence of chaotic behavior in spatiotemporal dynamics of complex natural systems does not require the effective number of dynamic variables needed to fully describe the system to be large [13–18], which disproves the statement that it is a priori impossible to get an in-depth understanding of the dynamics of complex objects. In reality, in spite of the visible chaoticity of time $V(t)$ ($t$ – time) and space $h(x)$ ($x$ – coordinate) variations in measured dynamic variables, there is some order related to scaling phenomena in the dynamics and structure of natural systems on different levels of their spatiotemporal hierarchy. It is the latter circumstance that leads to the low effective dimensionality of the resulting problems and opens up opportunities for developing practical methods to analyze the dynamics and structure of complex systems.

Many believed that the analysis of complex signals and extraction of the information hidden in them could be achieved using a methodology based on the theory of deterministic chaos [13,15,16]. It was suggested that a single measured variable can adequately describe the general dynamics of a complex system, and that the topology of a general attractor could be fully recreated by observing the dynamics of this single variable $V(t)$ [13–16]. The analysis based on $V(t)$ and variables $V(t+n\tau)$, where $n = 1, 2, …M$, introduces the concept of a multidimensional phase space of dimension $M$ in which the trajectory of the studied dynamic process is placed. The correlation dimension of an attractor, parameters of Poincaré sections, Lyapunov exponents, and Kolmogorov-Sinai entropy are taken as the dynamic parameters a collection of which can uniquely characterize the evolution of the system under study based on Takens' theorems, which are valid for stationary systems. However, this methodology did not succeed in the analysis of time series generated by real natural processes. In authors' opinion, this may be attributed to nonstationary evolution of real systems as well as simplified presentation of the information contained in the complex signals generated by real systems. The information units that are encountered in nature are more complex than simple bits. For example, the information unit in DNA structures is a codon, which represents a triplet of nucleotides. Hieroglyphics has its own information "clichés". Even in computer science, where bits are still the units of information, it is believed that the performance of computers can dramatically be increased only on the basis of radically different approaches to handling information, such as quantum computing, which manipulates matrix information units, qubits.

A significant progress in understanding the dynamics of complex systems was achieved due to numerous simulations using cellular automata models, which provided the grounds for introducing the phenomenon of Self-Organized Criticality (SOC) [17]. The idealized model simulating the formation of avalanches in sandpiles under the action of a constant external factor, such as pouring ("sprinkling") of sand on the top of the pile, in which avalanches of different sizes come down the slope, is considered to be the basic model with SOC behavior [19]. It was shown that the intermittent behavior of avalanches on all possible scales (numbers of sand particles per avalanche) led to power law distributions for the number of sand particles per avalanche. This allowed us to understand the well-known scale-invariant Guttenberg-Richter and Zipf-Parreto laws and explain the phenomenon of flicker noise, which are observed when the evolution is highly irregular and its behavior is non-Darwinian. It was demonstrated that the origins of such power laws, which imply the presence of long-range (infinite for flicker noise) time and space correlations in the systems despite the apparent chaoticity of the evolution and observed structures, are associated with the occurrence of complex (multiparticle, nonlinear) interactions, dissipation, and inertia.

The next milestone in understanding the phenomenon of complexity based on simulations using cellular automata and other discrete machines is the fundamental work of Stephen Wolfram [18]. He demonstrated that randomness and intermittency can be observed in discrete machines for which the underlying rules and initial conditions are fully deterministic. Deterministic and stochastic discrete machines were able to produce idealized models with complex behavior that show macroscopic (sometimes "visual") features close to those observed in natural systems. At the same time, these idealized models of complex processes do not allow one to directly interpret and analyze the chaotic signals generated by real natural systems.

The recent success of wavelet analysis in signal and image processing [20–22], especially image compression, raised a question on how much information about the evolution of complex systems and the structures it forms is stored in irregularities of measured dynamic variables. In fact, the wavelet transformation keeps just a small number of largest (in magnitude) decomposition coefficients and then uses only these coefficients to obtain the inverse transformation. This implies that the inverse transform uses only those time and space intervals for which the first derivative of the signal experiences the steepest changes, in other words, the intervals with the highest degree of irregular behavior.

Informativeness of signal irregularities at every level of the spatiotemporal hierarchy of the evolution under study became the basic hypothesis of Flicker-Noise Spectroscopy (FNS) [23–36], a general phenomenological framework for extracting the information stored in various complex signals. This framework allows one to develop algorithms that can be used to determine as many information parameters about the system state based on the analysis of its complex signals as is needed to solve the problems under study. This review provides the principles of FNS and illustrates its capabilities in solving some problems relevant to electrochemistry and surface science.

## 2. Fundamentals of Flicker-Noise Spectroscopy

Let us introduce the main concepts of FNS. For simplicity, we will consider the evolution dynamics of variable $V(t)$ that is experimentally measured on some time interval $T$. This variable may describe the dynamics of a physicochemical or natural process, fluctuations in one of the medical characteristics of a human organism, and so on. The goal is to extract information from recorded variations in $V(t)$, which may generally be chaotic. It is obvious that the features of every studied process are highly individual and determined by specific properties of the system as well as external forces and factors. The standard research problems are to find the main features of the dynamics, determine the role of external forces in the evolution, and predict possible changes in the system state. These problems require that a complete collection of digitized data $V(t)$ be characterized, and the physical essence of the information contained in a sequence of experimentally measured time series for dynamic variable $V(t)$ be understood. This can be accomplished only after the corresponding phenomenological parameters of the process, which do not depend on the individual features of the system and external forces affecting its evolution, are determined. At the same time, the introduced parameters of the dynamics should be related to physical processes in the system under study. In other words, we need an algorithm that can extract the necessary number of physically meaningful phenomenological parameters from the complete set of digitized time series for $V(t)$.

In FNS, these parameters are assumed to be related to the autocorrelation function, one of the basic concepts in statistical physics, which is defined as

$$\psi(\tau) = \langle V(t)V(t+\tau) \rangle_{T-\tau}, \tag{1}$$

where $\tau$ is the time lag parameter, and the angular brackets denote averaging:

$$\langle (...) \rangle_T = \frac{1}{T} \int_{-T/2}^{T/2} (...) \, dt. \tag{2}$$

Expression (1) implies that function $\psi(\tau)$ describes the relation between the values of dynamical variable $V(t)$ at larger and smaller values of the argument (we assume $\tau > 0$). The use of the averaging procedure over time interval $T$ means that all the characteristics extracted from the analysis of $\psi(\tau)$ should be considered as average ones for that interval. If $T$ is a subinterval of $T_{tot}$ ($T < T_{tot}$), then the value of function $\psi(\tau)$ may depend on the position of $T$ inside the larger interval $T_{tot}$. If there is no such dependence and $\psi(\tau)$ is determined only by the difference in the arguments on the right-hand side of Eq. (1), then the evolution process is called "stationary".

The averaging procedure can be written for the discrete case if one divides the interval $T$ into $N$ small intervals $\Delta t$ such that $N = \lfloor T/\Delta t \rfloor$. Introducing $n_\tau = \lfloor \tau/\Delta t \rfloor$, we obtain

$$\psi(n_\tau) = \frac{1}{N - n_\tau} \sum_{k=1}^{N-n_\tau} V(k)V(k+n_\tau). \tag{3}$$

This implies that the value of autocorrelator $\psi(n_\tau)$ at every value of argument $n_\tau$ is the average of product $V(k)V(k+n_\tau)$ over all values $k$ inside the interval $[0, N-n_\tau]$.

To extract the information contained in $\psi(\tau)$, it is convenient to analyze some transforms ("projections") of this function, for example, "power spectrum" $S(f)$:

$$S(f) = \left| \int_{-T_M/2}^{T_M/2} \left[ \psi(t_1) - \langle V(t) \rangle_T^2 \right] \exp(-2\pi i \, f t_1) \, dt_1 \right|, \tag{4}$$

or in discrete form:

$$S(q) = \left| \sum_{m=0}^{M-1} \left[ \psi(m) - \left( \frac{1}{N} \sum_{k=1}^{N} V(k) \right)^2 \right] \exp\left( -\frac{2\pi i \, qm}{M} \right) \right|, \tag{5}$$

where $f$ is the frequency, $q = fT_M$, $T_M < T$, $M = \left\lfloor \dfrac{T_M}{T} N \right\rfloor$ is the number of points on the frequency axis, and $S(q) = \Delta t \times S(f)$. As the autocorrelator is symmetric, $\psi(m) = \psi(M-m)$. This particular transform was chosen because $S(f)$ is most effective in separating out the resonances (main components of the signal) of the analyzed functions, which are represented as a set of $n$ peaks characterized by positions $f_{0i}$ and "half-widths" $\gamma_i$ ($i = 1, 2, \ldots, n$).

The resonance contribution $S_r(f)$ to the overall power spectrum $S(f)$ can be extracted by rewriting the latter as

$$S(f) = S_c(f) + S_r(f), \tag{6}$$

where $S_c(f)$ is the continuous power-spectrum component associated with the "chaotic" component of dynamic variable $V(t)$ [13,14]. Additive representation (6) is justified by the fact that the resonant components of dynamic variable $V(t)$ change rather "smoothly" [28]. Let us note that the parameterization of $S_r(f)$ by finding the positions, "half-widths", and partial weights $A_i$ of the fixed resonances can be done rather easily. At the same time, the parameterization of $S_c(f)$ is a much more difficult task.

To solve the latter problem, we also consider difference moments ("transient structural functions") $\Phi^{(p)}(\tau)$ of different orders $p$ ($p = 2, 3, \ldots$):

$$\Phi^{(p)}(\tau) = \left\langle [V(t) - V(t+\tau)]^p \right\rangle_{T-\tau}, \tag{7}$$

or in discrete form:

$$\Phi^{(p)}(n_\tau) = \frac{1}{N - n_\tau} \sum_{k=1}^{N-n_\tau} [V(k) - V(k+n_\tau)]^p. \tag{8}$$

When $p \geq 3$, it is more convenient to use dimensionless "transient quasi-cumulants" $\mu^{(p)}(\tau)$ instead of $\Phi^{(p)}(\tau)$:

$$\mu^{(p)}(\tau) = \frac{\Phi^{(p)}(\tau)}{\left[\Phi^{(2)}(\tau)\right]^{p/2}}. \tag{9}$$

It is obvious that when $p = 2$, we have

$$\Phi^{(2)}(\tau) = 2[\psi(0) - \psi(\tau)]; \tag{10}$$

that is, $\Phi^{(2)}(\tau)$ linearly depends on $\psi(\tau)$. This function can be regarded as another "projection" of $\psi(\tau)$ allowing one to disclose some additional information stored in the autocorrelator, which complements the information of power spectrum $S(f)$ given by Eq. (4). Function (10) will also be written as a linear combination of chaotic $\Phi_c^{(p)}(\tau)$ and resonant $\Phi_r^{(p)}(\tau)$ components [28]:

$$\Phi^{(2)}(\tau) = \Phi_c^{(2)}(\tau) + \Phi_r^{(2)}(\tau). \tag{11}$$

The main problem in parameterizing the chaotic components can be illustrated by considering the case in which the resonance contributions to $V(t)$ are absent and function $V(t)$ is completely "chaotic" (Figs. 1a,b). Here, the expressions for functions $\Phi_c^{(2)}(\tau)$ and $S_c(f)$ obtained for "stationary" processes can be written as

$$\Phi_c^{(2)}(\tau) \to \begin{cases} \tau^{2H_1}, & \text{if } \tau < T_1 \\ 2\sigma^2, & \text{if } \tau > T_1 \end{cases}, \quad \sigma^2 \equiv \langle V^2 \rangle - \langle V \rangle^2, \tag{12}$$

where $T_1$ is the "correlation time" determined from $\Phi_c^{(2)}(\tau)$, the parameter $H_1$ is the Hurst constant, $\sigma$ is the variance of measured values of $V(t)$; and

$$S_c(f) \to \begin{cases} 1/f^n, & \text{if } f > 1/T_0 \\ S_c(0), & \text{if } f < 1/T_0 \end{cases}, \tag{13}$$

where $T_0$ is the "correlation time" determined from $S_c(f)$, $n$ and $S_c(0)$ are parameters.

Parameters $T_1$, $H_1$, $\sigma$, $T_0$, $n$ and $S_c(0)$ may be given a specific physical interpretation and can be considered as phenomenological parameters that characterize the chaotic component of signal $V(t)$. Indeed, it follows from Eq. (12) and Fig. 1a that parameter $T_1$ determines the characteristic time interval during which the measured value of $V(t)$ gets "forgotten". In other words, the values of dynamic variables stop correlating when their arguments are $T_1$ apart from each other. It is clear that $\Phi_c^{(2)}(\tau) = 0$ when $\tau = 0$ and $\Phi_c^{(2)}(\tau) \to 2\sigma^2$ when $\tau > T_1$. Hence, to obtain reliable estimates of the variance, one needs to calculate it for the intervals exceeding $T_1$. In this case, $H_1$ describes the "law" of correlation loss for the values of $V(t)$ measured at different time moments. Dimensionless parameter $n$ describes the correlation loss with frequency decline to the value of $1/T_0$ (see Fig. 1b). At the same time, the physical interpretation of parameter $T_0$ is not completely clear: it can be a correlation loss in the values of $V(t)$, as it is in the case of $T_1$, or some other characteristic of dynamic variable $V(t)$.

As both $S_c(f)$ and $\Phi_c^{(2)}(\tau)$ are expressed in terms of the autocorrelation function, one may think that the corresponding parameters $T_1$ and $T_0$, $H_1$ and $n$ are related, and thus the number of independent parameters involved in $\Phi_c^{(2)}(\tau)$ and $S_c(f)$ can be reduced. However, our experience in the analysis of time series $V(t)$ corresponding to the evolution of real processes shows that the expected relations

$$T_1 = T_0, \quad 2H_1 + 1 = n \tag{14}$$

do not hold true [27,28]. Let us note that the second expression in Eq. (14) is virtually the well-known "fractal self-similarity" relation for evolving processes. The lack of relation between the information parameters involved in $S_c(f)$ and $\Phi_c^{(2)}(\tau)$ may be attributed either to the "nonstationarity" of real processes $V(t)$ or finiteness of the averaging intervals $T$. This poses the question of which of the functions, $S(f)$ or $\Phi^{(2)}(\tau)$, should be used as the basic function that can adequately parameterize real processes which are definitely analyzed on finite time intervals $T$ and may be "nonstationary". Prior to FNS, this question was open, and the analysis of various chaotic signals was primarily based on $S(f)$ rather than $\Phi^{(2)}(\tau)$.

This problem was resolved after the concept of information contained in various complex signals was generalized. According to FNS, the information contents of $S_c(f)$ and $\Phi_c^{(p)}(\tau)$ are different. Thus, in order to determine all system parameters that are required to solve practical problems, one needs to analyze both functions.

The basic idea of FNS is to treat the correlation links present in sequences of different irregularities, such as spikes, "jumps", discontinuities in derivatives of different orders, on all levels of the spatiotemporal hierarchy of the system under study as the main information carriers. It is further assumed that according to the SOC paradigm [17], the chaotic dynamics of real processes is associated with *intermittency*, consecutive alternation of rapid changes in the values of dynamic variables on small time intervals with small variations of the values on longer time intervals. Such intermittency occurs on every hierarchical level of the system evolution.

The functions $\Phi_c^{(p)}(\tau)$ are formed exclusively by "jumps" of the dynamic variable while $S_c(f)$ is formed by both "spikes" and "jumps" on every level of the hierarchy [23]. To illustrate this statement, consider the process of one-dimensional "random walk" with small "kinematic viscosity" $v$ (Fig. 2). The small value of $v$ implies that when the signal changes from position $V_i$ to $V_{i+1}$, which are $|V_{i+1} - V_i|$ apart (in value) from each other, the system first overleaps ("overreacts") due to inertia and then "relaxes". We assume that the relaxation time is small compared to the residence time in a "fluctuation" position. It is obvious that when the number of walks is large, the functions $\Phi^{(p)}(\tau)$ will not depend on the values of "inertial skipovers" of the system, but will be determined only by the algebraic sum of walk "jumps". At the same time, the functions $S(f)$, which characterize the "energy side" of the process, will be determined by both spikes and jumps.

It should be underlined that such separation of information stored in various irregularities is attributed to the intermittent character of the evolution dynamics. Indeed, the information contents of $S_c(f)$ and $\Phi_c^{(2)}(\tau)$ coincide if there is no intermittence, as happens in the case of completely "irregular" dynamics of the Weierstrass – Mandelbrot (WM) function. Let us demonstrate it.

The real part of WM function is written as [37,38]:

$$F_{WM}(t) = \sum_{n=-\infty}^{\infty} \frac{1-\cos b^n t}{b^{(2-D)n}} \qquad (b>1,\ 1<D<2).$$

Though continuous, this function cannot be differentiated at any point.

Autocorrelator $\psi_{WM}$, transient difference moment of second order $\Phi_{WM}^{(2)}(\tau)$, and power spectrum $S_{WM}(f)$ for WM are expressed as [37]:

$$\psi_{WM}(\tau) = \langle F_{WM}(t) F_{WM}(t+\tau) \rangle = \frac{1}{2} \sum_{n=-\infty}^{\infty} \frac{\cos b^n \tau}{b^{2(2-D)n}},$$

$$\Phi_{WM}^{(2)}(\tau) = \langle [F_{WM}(t) - F_{WM}(t+\tau)]^2 \rangle = \sum_{n=-\infty}^{\infty} \frac{1-\cos b^n \tau}{b^{2(2-D)n}},$$

$$S_{WM}(f) = \left| 2 \int_0^{\infty} F_{WM}(\tau) \cos(2\pi f \tau) d\tau \right|^2 = \frac{1}{4} \sum_{n=-\infty}^{\infty} \frac{\delta(2\pi f - b^n)}{b^{2(2-D)n}} \xrightarrow[2\pi(b-1)f \ll 1]{} \frac{1}{4 \ln b \, (2\pi f)^{5-2D}}.$$

In this case, the constant ($\tau$-independent) term in the expression for $\psi_{WM}(\tau)$ was discarded. The corresponding term in the expression for power spectrum $S_{WM}(f)$, which characterizes the null frequency, was also discarded.

It is easy to show that functions $\Phi_{WM}^{(2)}(\tau)$ and $S_{WM}(f)$ can be expressed in terms of each other:

$$\psi_{WM}(\tau) = 2 \int_0^{\infty} S_{WM}(f) \cos(2\pi f \tau) = \frac{1}{2} \sum_{n=-\infty}^{\infty} \frac{\cos b^n \tau}{b^{2(2-D)n}},$$

$$\Phi_{WM}^{(2)}(\tau) = 2 [\psi_{WM}(0) - \psi_{WM}(\tau)] = \sum_{n=-\infty}^{\infty} \frac{1-\cos b^n \tau}{b^{2(2-D)n}},$$

$$S_{WM}(f) = 2 \int_0^{\infty} \psi_{WM}(\tau) \cos(2\pi f \tau) d\tau = \frac{1}{4} \sum_{n=-\infty}^{\infty} \frac{\delta(2\pi f - b^n)}{b^{2(2-D)n}}.$$

Hence, the information contents of $\Phi_{WM}^{(2)}(\tau)$ and $S_{WM}(f)$ are the same despite the "chaotic" nature of WM function.

It is important to clarify that when we talk about the contributions of various irregularities to $S_c(f)$ and $\Phi_c^{(p)}(\tau)$, we do not mean the "jumps", "spikes", and derivative discontinuities that are visually discernible in the measured chaotic variations of $V(t)$. Because if we increase the discretization frequency of the signal, which plays the role of resolution in temporal analysis, each of the irregularities observed at a smaller discretization frequency becomes a complex structure with its own "spikes" and "jumps". Formally speaking, it is possible to introduce an infinite number of discretization levels that are mapped onto an infinite number of levels in the evolution hierarchy. Hence, the "irregularity" observed at a specific discretization frequency is formed by an infinite set of irregularities corresponding to smaller time scales of the hierarchy.

The goal of FNS, which is a phenomenological approach to the parameterization of complex chaotic signals, is to determine physically meaningful parameters of the dynamics, the number of which depends on the specifics of the problem under study. FNS is a deductive theory that is based on the idealized concepts of Dirac δ-function (spike) and Heaviside θ-function (jump). These irregularities are introduced as the carriers of information about the system dynamics on every level of the infinity of hierarchical evolution levels. The information parameters of the evolution dynamics on every $i$-th level can be extracted from functions $S_{ci}(f)$ and $\Phi_{ci}^{(p)}(\tau)$, which are calculated based on the sequences of irregularities corresponding to this level. It is at this stage that one observes the information difference in functions $S_{ci}(f)$ and $\Phi_{ci}^{(p)}(\tau)$, which are determined by jumps and spikes and jumps alone, respectively. This can be easily understood using the well-known result of Schuster [13], who showed that the power spectrum, which corresponds to a correlated sequence of δ-functions, at low frequencies, when the averaging is carried out for time intervals much larger than the characteristic time interval between the adjacent δ-functions, may be continuous. It was mentioned above that this is a feature of chaotic signals [14]. At the same time, the structural function $\Phi_{ci}^{(p)}(\tau)$ is equal to zero because the domain set of a δ-function sequence is a set of measure zero [39]. Function $\Phi_{ci}^{(p)}(\tau)$ takes a nonzero value at a specific level of the hierarchy when θ-functions are inserted in the sequence of δ-functions.

To analyze the functions $S_c(f)$ and $\Phi_c^{(p)}(\tau)$ corresponding to real complex systems, it is necessary to know the integral contribution from all levels of the evolution hierarchy at a specific discretization frequency. The situation is very simple in the case of stationary evolution, when all levels of the hierarchy are described by the same functions $S_{ci}(f)$ and $\Phi_{ci}^{(p)}(\tau)$. However, the real evolution processes are generally nonstationary. At the same time, one may identify the time scales ("windows") where the evolution may be considered as "quasi-stationary". Such introduction of nonstationarity on "small" time intervals has much in common with the application of Short Time Fourier Transform (STFT) to the analysis of nonstationary signals; i.e., signals with time varying spectra. Indeed, the transition from the usual Fourier Transform (FT), which does not contain any information about the time localization of spectral components, to STFT is effected by introducing the concept of a "window" inside which the signal is assumed to be stationary [40]. A similar procedure will be used in Chapter 4 to derive the FNS criteria of nonstationarity.

The information "passport" characteristics that are determined by fitting the derived expressions to the experimental variations of $S_c(f)$ and $\Phi_c^{(p)}(\tau)$ are interpreted as the correlation times and parameters, which describe the rate of "memory loss" on these correlation time intervals for different irregularities. In the case of "stationary" evolution, these parameters have the same values on every level of the hierarchy. Hence, in contrast to the geometric self-similarity on different scales in the theory of fractals and multifractals, FNS introduces the dynamic self-similarity on all levels of the hierarchy. The latter self-similarity is multiparametric as $S_c(f)$ and $\Phi_c^{(p)}(\tau)$ are generally characterized by a set of parameters.

In the next three chapters, we will present main FNS expressions that are used in the analysis of various signals. The substantiation and derivation of these expressions are presented elsewhere [26].

## 3. FNS expressions for stationary processes

Let us write the interpolation expressions for chaotic components $\Phi_c^{(2)}(\tau)$ and $S_c(f)$ for stationary processes. The parameters characterizing the dynamic correlations on every level of the evolution hierarchy are assumed to be the same. Consider the simplest case, in which there is only one characteristic scale in the sequences of "spikes" and "jumps". Then [23]:

$$\Phi_c^{(2)}(\tau) \approx 2\sigma^2 \cdot \left[1 - \Gamma^{-1}(H_1) \cdot \Gamma(H_1, \tau/T_1)\right]^2, \tag{15}$$

$$\Gamma(s,x) = \int_x^\infty \exp(-t) \cdot t^{s-1} dt, \quad \Gamma(s) = \Gamma(s,0),$$

where $\Gamma(s)$ and $\Gamma(s, x)$ are the complete and incomplete gamma functions ($x \geq 0$ and $s > 0$), respectively. As in Eq. (12), $\sigma$ in Eq. (15) is the variance of measured values of $V(t)$; $H_1$ is the Hurst constant, which describes the rate of "forgetting" by the dynamic variable of its values on time intervals that are less than the correlation time $T_1$. In this case, $T_1$ may be interpreted as the correlation time for the "jumps" in a chaotically varying $V(t)$.

For particular cases, we have:

$$\Phi_c^{(2)}(\tau) = 2\Gamma^{-2}(1+H_1) \cdot \sigma^2 \left(\frac{\tau}{T_1}\right)^{2H_1}, \quad \text{if } \frac{\tau}{T_1} \ll 1 \tag{16}$$

$$\Phi_c^{(2)}(\tau) = 2\sigma^2 \left[1 - \Gamma^{-1}(H_1) \cdot \left(\frac{\tau}{T_1}\right)^{H_1-1} \exp\left(-\frac{\tau}{T_1}\right)\right]^2, \quad \text{if } \frac{\tau}{T_1} \gg 1. \tag{17}$$

The interpolating function for power spectrum component $S_{cS}(f)$ formed by the "jumps" is written as [23]:

$$S_{cS}(f) \approx \frac{S_{cS}(0)}{1+(2\pi f T_0)^{n_0}}. \tag{18}$$

Here, $S_{cS}(0)$ is the parameter characterizing the low-frequency limit of $S_{cS}(f)$, $n_0$ describes the degree of correlation loss in a sequence of the "spikes" on time interval $T_0$.

The interpolating function for power spectrum component $S_{cR}(f)$ formed by the "spikes" is written as [23]:

$$S_{cR}(f) \approx \frac{S_{cR}(0)}{1+(2\pi f T_1)^{2H_1+1}}, \tag{19}$$

where

$$S_{cR}(0) = 4\sigma^2 T_1 H_1 \cdot \left\{1 - \frac{1}{2H_1\Gamma^2(H_1)} \int_0^\infty \Gamma^2(H_1,\xi)d\xi\right\}. \tag{20}$$

Although the contributions to the overall power spectrum $S_c(f)$ given by Eqs. (18) and (19) are similar, the parameters in those equations are different; that is, $S_{cS}(0) \neq S_{cR}(0)$, $T_1 \neq T_0$, and $2H_1 + 1 \neq n_0$. This implies that the parameters in the expressions for the power spectrum and structural function of second order generally have different information contents when one analyzes experimental time series $V(t)$.

The interpolating function for the overall power spectrum can be expressed as:

$$S_c(f) \approx \frac{S_c(0)}{1+(2\pi f T_{01})^n}, \tag{21}$$

where $S_c(0)$, $T_{01}$, and $n$ are phenomenological parameters calculated by fitting the interpolating function (21) to the power spectrum for experimental time series.

As was noted above, complex signals usually contain both "chaotic" (formed by irregularities) and resonant (system specific) components. Therefore, our goal is to extract all parameters for both the chaotic component and resonances from time series $V(t)$.

The overall signal $V(t)$ will be represented as a linear superposition of the "high-frequency" chaotic component $V_c(t)$ and "slow-varying" resonant component $V_r(t)$; that is,

$V(t) = V_c(t) + V_r(t)$.

If we consider large time intervals ($T >> \max\{T_0, T_1\}$), and the autocorrelator $\psi_r(\tau) = <V_r(t)V_r(t+\tau)>$ averaged over time interval $T$ depends only on the time lag parameter $\tau$, then the functions $S_r(f)$ and $\Phi^{(2)}_r(\tau)$, which correspond to $V_r(t)$, become interrelated:

$$S_r(f) = \int_0^\infty \cos(2\pi f \tau)\left[\Phi^{(2)}_r(\infty) - \Phi^{(2)}_r(\tau)\right]d\tau. \tag{22}$$

This expression can be derived from (1) and (10) in view of $\Phi^{(2)}_r(\infty) = 2[\psi_r(0) - <V_r(t)>^2]$.

Consider the model case in which the resonance contribution $S_r(f)$ to the overall power spectrum is of Lorenz type with parameters $f_0$ ("resonance position"), $\gamma$ («half-width»), and $A$ («intensity»):

$$S_r(f) = A \cdot \left[\frac{1}{(f-f_0)^2+(\gamma/2\pi)^2} + \frac{1}{(f+f_0)^2+(\gamma/2\pi)^2}\right]. \tag{23}$$

The corresponding expression for autocorrelation function $\psi_r(\tau)$ takes the form:

$$\psi_r(\tau) = 2\int_0^\infty S_r(f)\cos(2\pi f \tau)df = 4A\gamma^{-1}\pi^2\exp(-\gamma\tau)\cos(2\pi f_0 \tau). \tag{24}$$

Thus, for $\Phi^{(2)}_r(\tau)$ we obtain

$$\Phi^{(2)}_r(\tau) = 2[\psi_r(0) - \psi_s(\tau)] = 8A\gamma^{-1}\pi^2[1 - \exp(-\gamma\tau)\cos(2\pi f_0 \tau)]. \tag{25}$$

The signal under study may generally include several resonances. In this case, the expressions for $S_r(f)$ and $\Phi^{(2)}_r(\tau)$ are written as

$$S_r(f) = \sum_{i=1}^m A_i\left[\frac{1}{(f-f_{0i})^2+(\gamma/2\pi)^2} + \frac{1}{(f+f_{0i})^2+(f/2\pi)^2}\right], \tag{26}$$

$$\Phi_r^{(2)}(\tau) = \sum_{i=1}^{m} \alpha_i [1 - \exp(-\gamma_i \tau) \cos(2\pi f_{0i} \tau)]; \quad \alpha_i \equiv 8\pi^2 A_i \gamma_i^{-1}, \tag{27}$$

where $f_{0i}$, $\gamma_i$, and $A_i$ ($i = 1, 2, \ldots, m$) are the parameters for the $i$-th resonance.

Expressions (15)-(20) along with (26) and (27) can be used to describe the variations (4), (5), (7), (8) and (10) calculated from the experimental time series $V(t)$. This procedure can be used to determine the values of parameters both for the chaotic ($n_0$, $T_0$, $S_{cS}(0)$, $H_1$, $T_1$, $\sigma$) and resonant ($f_{0i}$, $\gamma_i$ and $A_i$) signal components.

To conclude the chapter, we list below *the algorithm for calculating the parameters that characterize the dynamics and/or structural features of open complex (physicochemical, natural) systems*:

- Consider the simplest case in which the difference moment of second order $\Phi^{(2)}(\tau)$ and power spectrum $S(f)$ are good enough to determine the necessary set of parameters.
- STEP 1. Using the experimental data, evaluate the difference moment of second order $\Phi^{(2)}(\tau)$ with Eq. (10), where autocorrelator $\psi(\tau)$ is given by Eq. (3). Evaluate power spectrum $S(f)$ with Eq. (5).
- STEP 2. We assume that power spectrum $S_r(f)$ near the resonances can be approximated by Eq. (26), which allows us to determine parameters $f_{0i}$, $\gamma_i$ and $A_i$ ($i = 1, 2, \ldots, m$) for the main resonances. The optimal number of the resonances is chosen based on the steps that follow.
- STEP 3. If the experimental function $S(f)$ includes a continuous (chaotic) component $S_c(f)$, the latter can be subtracted from $S(f)$ using interpolation expression (21). This allows us to determine parameters $S_c(0)$, $T_{01}$, and $n$.
- STEP 4. Using the determined values of $f_{0i}$, $\gamma_i$, and $A_i$ ($i = 1, 2, \ldots, m$), evaluate $\Phi^{(2)}_r(\tau)$ with Eq. (27). Add the latter to $\Phi^{(2)}_c(\tau)$ calculated with Eq. (15). Find unknown parameters $H_1$, $T_1$, $\sigma$ by fitting this sum to the experimental function (10).
- Steps 2-4 are iterated until the error of the calculated function compared to the experimental dependence (10) reaches some acceptable value (using the method of least squares). Once the minimization is completed, the optimal values of parameters ($n_0$, $T_0$, $S_{cS}(0)$, $H_1$, $T_1$, $\sigma$) and ($f_{0i}$, $\gamma_i$, $A_i$) are known.

## 4. Dynamics of nonstationary processes: precursors of catastrophic system state changes

To analyze the effects of nonstationarity in real processes, it is suggested to study the dynamics of changes in $S(f)$ and $\Phi^{(p)}(\tau)$ for consecutive "window" intervals $[t_k, t_k+T]$, where $k = 0, 1, 2, 3, \ldots$ and $t_k = k\Delta T$, that are shifted within the total time interval $T_{\text{tot}}$ of experimental time series ($t_k+T < T_{\text{tot}}$). The time intervals $T$ and $\Delta T$ are chosen based on the physical understanding of the problem in view of the suggested characteristic time of the process, which is the most important parameter of the system evolution. If there are secondary processes with characteristic times $\tau_i$, which play an insignificant role in the main nonstationary process of structure reconfiguration, then the interval $T$ should be chosen so that $\tau_i \ll T$ be true.

It seems reasonable to relate the phenomenon of "precursor" occurrence with abrupt changes in functions $S(f)$ and $\Phi^{(p)}(\tau)$ when the upper bound of the interval $[t_k, t_k+T]$ approaches the time moment $t_c$ of a catastrophic event accompanied by total system reconfiguration on all space scales. It is obvious that one can talk about a "precursor" only if time $t_k$ is at least $\Delta T$ less than time $t_c$; that is, $\Delta T_{sn} = t_c - t_k \geq \Delta T$ assuming $\Delta T_{sn} \ll T_{tot}$.

The analysis of experimental chaotic series often requires the original data to be smoothed. There are several ways to filter the digitized signal with extracting the low-frequency component: smoothing polynomials, wavelets, and so on. The "relaxation" procedure proposed in [28] uses the analogy with a finite-difference solution of the diffusion equation, which allows one to split the original signal into low-frequency $V_R(t)$ and high-frequency $V_F(t)$ components. The iterative procedure which finds the new values of the signal on every "relaxation" step using its values for the previous step allows one to determine the low-frequency component $V_R$. The high-frequency component $V_F$ is obtained by subtracting $V_R$ from the original signal. Basically, this smoothing algorithm progressively reduces the local gradients of the "concentration" variables, causing the points in every triplet to come closer to each other. Such splitting of the original signal $V(t)$ into $V_R(t)$ and $V_F(t)$ makes it possible to evaluate functions $S(f)$ and $\Phi^{(p)}(\tau)$, the expressions for which are derived above, for each of the three functions $V_J(t)$ (J = R, F, or G), where index G corresponds to the cases when smoothing is not needed.

For example, the determination of the precursors of catastrophic events calculated using difference moments $\Phi_J^{(p)}(\tau)$ given by Eqs. (7) and (8) uses both the high-frequency and low-frequency components $V_J(t)$. Note that functions $\Phi_J^{(p)}(\tau)$ can be reliably evaluated only on the $\tau$ interval of $[0, \alpha T]$, which is less than a half of the averaging interval $T$; i.e., $\alpha < 0.5$. Presently, the most illustrative precursor of catastrophic events is the spikes in nonstationarity factors, which are defined by dimensionless relations involving the difference moment $\Phi^{(2)}(\tau)$ [28, 41]:

$$C_J(t_{k+1}+T) = \frac{Q_{k+1}^J - Q_k^J}{\Delta T} \times \frac{2T}{Q_{k+1}^J + Q_k^J}, \tag{28}$$

where

$$Q_k^J = \frac{1}{\alpha T}\int_0^{\alpha T} \left[\Phi_J^{(2)}(\tau)\right]_k d\tau,$$

or in discrete form ($b = \lfloor \Delta T / \Delta t \rfloor$, $N_1 = \lfloor \alpha N \rfloor$):

$$Q_k^J = \frac{1}{N_1} \sum_{n_\tau=1}^{N_1} \left[ \Phi_J^{(p)}(n_\tau) \right]_k = \frac{1}{N_1} \sum_{n_\tau=1}^{N_1} \frac{1}{N-n_\tau} \sum_{m=1+kb}^{N-n_\tau+kb} [V(m)-V(m+n_\tau)]^p .$$

Here, J indicates which function $V_J(t)$ (J = R, F or G) is used to evaluate $\Phi_J^{(2)}(\tau)$.

In general, FNS precursors provide the "nonstationarity measure" of the process under study when the averaging interval $T$ is shifted by $\Delta T$ along the time axis, which is of particular interest when the upper bound of the time interval $[t_k, t_k+T]$ approaches the time $t_c$ of a catastrophic event.

To conclude the chapter, we list below *the algorithm for finding the precursors of abrupt changes in the state of various open dissipative systems based on the a priori information about their dynamics*:

- STEP 1. Select time intervals $T$ and $\Delta T$ from physical considerations.
- STEP 2. For "sliding" windows of averaging, evaluate $\Phi^{(2)}(\tau)$ with Eq. (10). Then, evaluate non-stationary factor $C_J(t_{k+1}+T)$ with Eq. (28). The inequality $\tau \leq 0.5T$ must be true.

## 5. Correlation links in the dynamics of distributed systems

FNS opens up new opportunities for the analysis of different (mass, electric, magnetic) flows in distributed systems with the following two characteristics: (1) The system size is larger than the space scale of noticeable variations in the dynamic variables. (2) The evolution is "perplexed" due to complex nonlinear interactions between the components of the system. The information about the dynamics of correlation links in variables $V_i(t)$, measured at different points $i$, can be extracted by analyzing the temporal links of various correlators [24, 42].

In this review, we will limit our attention to the simplest "two-point" correlators characterizing the links between $V_i(t)$ and $V_j(t)$ [24]:

$$q_{ij}(\tau; \theta_{ij}) = \left\langle \left[ \frac{V_i(t)-V_i(t+\tau)}{\sqrt{\Phi_i^{(2)}(\tau)}} \right] \left[ \frac{V_j(t+\theta_{ij})-V_j(t+\theta_{ij}+\tau)}{\sqrt{\Phi_j^{(2)}(\tau)}} \right] \right\rangle_{T-\tau-|\theta_{ij}|}, \quad (29)$$

or in discrete form, introducing $n_\theta = \lfloor \theta_{ij}/\Delta t \rfloor$:

$$q_{ij}(n_\tau; n_\theta) = \frac{N-n_\tau}{N-n_\tau-|n_\theta|} \frac{\sum_{k=U[-n_\theta]|n_\theta|+1}^{N-n_\tau-U[n_\theta]|n_\theta|} [V_i(k)-V_i(k+n_\tau)][V_j(k+n_\theta)-V_j(k+n_\theta+n_\tau)]}{\sqrt{\sum_{k=1}^{N-n_\tau}[V_i(k)-V_i(k+n_\tau)]^2} \sqrt{\sum_{k=1}^{N-n_\tau}[V_j(k)-V_j(k+n_\tau)]^2}}, \quad (30)$$

$$U[x] = \begin{cases} 1, & x \geq 0; \\ 0, & x < 0. \end{cases}$$

Here, $\tau$ is the "lag time"; $\theta_{ij}$ is the "time shift" parameter.

The dependence of correlator $q_{ij}(\tau, \theta_{ij})$ on $\theta_{ij}$ describes the cause-and-effect relation ("flow direction") between signals $V_i(t)$ and $V_j(t)$. When $\theta_{ij} > 0$, the flow moves from point $i$ to point $j$, when $\theta_{ij} < 0$, from $j$ to $i$. When the distance between points $i$ and $j$ is fixed, the value of $\theta_{ij}$ can be used to estimate the rate of information transfer between these two points. The dependence of the value and magnitude of correlator $q_{ij}(\tau, \theta_{ij})$ on $\tau$ and $\theta_{ij}$ can be used to analyze the flow dynamics with signals $V_i(t)$ and $V_j(t)$ changing in phase ($q_{ij} > 0$) and in antiphase ($q_{ij} < 0$).

To conclude the chapter, we list below *the algorithm for determining the flow dynamics in distributed systems based on the analysis of dynamic correlations in chaotic signals that are simultaneously measured at different points in space:*

- STEP 1. Select time interval $T$ from physical considerations.
- STEP 2. Evaluate $q_{ij}(\tau, \theta_{ij})$ with Eq. (30) for $\tau < 0.5\,T$ and $\theta_{ij} < \tau$.

## 6. Applications of FNS methodology

### 6.1. General remarks

FNS methodology may be applied to the extraction of information from experimental time series as well as spatial series. In the latter case, when the dynamic variable (magnitude of surface roughness, space distribution in concentrations of chemical substances, codon variations in a genome, etc.) varies chaotically along the configuration coordinate $x$ [29–31], one needs to make the following substitutions: $t \to x$ and $f \to f_x \equiv k/2\pi$, where the dimension of "space frequency" $f_x$ and wave number $k$ is $[x]^{-1}$.

If one analyzes time series $V(t)$, which characterize the fluctuation dynamics of a system as a whole, the dimensions of the system should not exceed its characteristic correlation length by orders of magnitude. Otherwise, the sequences of correlated fluctuations will take up only a small part of the overall system space, leading to a large number of such fluctuation sequences in the total volume. As all of these sequences are independent of each other (they are much further apart than their correlation length), the fluctuations averaged over the total volume will be described by a frequency-independent power spectrum.

FNS has been used to study the dynamics of various processes [26–35]: voltage fluctuations in electrochemical systems; fluctuations of velocity components in turbulent flows; fluctuation dynamics of solar activity, and so on. It has also been used to parameterize (in terms of "passport" parameters and/or patterns) surface structures studied with scanning probe microscopy, and chaotic segments of infrared or Raman spectra for complex compounds, which are often referred to as the "fingerprint" regions. Some of the results of these studies are presented below.

*6.2. Determination of parameters characterizing the dynamics or structural features of open complex systems*

*6.2.1. FNS parameterization of AFM images for surfaces of LiF single crystals*

Let us show how FNS can be used to characterize surface structures with "passport" parameters. Consider an example of surface profiles obtained by Atomic Force Microscopy (AFM), which is often used in metrology of surfaces at nano- and microscales [27]. AFM images represent digitized surface "roughness profiles" $h(x)$ along coordinate $x$ in the interval of $0 \leq x \leq L$ within the visible image window. To analyze these "spatial" signals with FNS, it is necessary to make substitutions $t \to x$ and $f \to f_x \equiv k/2\pi$, as well as $\tau \to \Delta$, $T_1 \to L_1$, $T_{01} \to L_{01}$ in Eqs. (3), (10), (15), (18)-(21), (26), and (27), which are used to determine the parameters. In Eq. (15), $\sigma$ is the variance of roughness values, $H_1$ is the Hurst constant, which describes the correlation "forgetting" rate for the "jump" irregularities in chaotic roughness profiles $h(x)$ on space intervals less then the correlation length $L_1$.

The traditional surface parameters are the arithmetic mean $<h>$ of digitized roughness profiles and their variance $\sigma$. However, when one analyzes real surfaces measured with AFM, these parameters cannot completely characterize the surface and thus cannot be used as the basic metrological parameters. Even when another parameter, fractal dimension $D$ [38,43,44], is introduced as the third "passport" characteristic, the surface metrology problem remains unsolved. It is obvious a priori that the individuality of a real rough surface (by the way, any real surface is "rough" at nanoscales) cannot be fully described with only three parameters.

Our goal is to extract information from AFM profiles $h(x)$, which may generally be chaotic. It is evident that the features of these profiles depend on physicochemical characteristics of the specimen material, as well as on the conditions of surface formation, modification, operation, etc. It is this type of relations that usually are of most interest to researchers and engineers. To determine these relations, the complete collection of digitized AFM data $h(x)$ must be characterized. This can be accomplished only after the corresponding phenomenological parameters, which do not depend on the technology of surface formation and physicochemical properties of the initial material, are determined. At the same time, the introduced surface parameters should be related to the physical properties of the measured profiles. In other words, we need an algorithm that can extract the necessary number of physically meaningful phenomenological parameters from the complete set of digitized AFM data for $h(x)$.

In [29,30], FNS was used to analyze the AFM images of structures on surfaces that are sufficiently homogeneous (without noticeable structure patterns). The AFM images were obtained for surfaces of lithium fluoride single crystals grown from a melt by the Strogbarger method. The crystals were cleaved along the (100) plane. A part of the specimens were treated with hydrogen in an autoclave at 500°C and 20 MPa for one week and then subjected to surface dissolution for one hour at different temperatures. All the primary data were obtained using an AFM Model Solver P-47 manufactured by the NT-MTD company (Zelenograd, Russia), which was operated in an intermittent contact (tapping) mode. The dimensions of the surface areas under study were varied from 7 to 50 μm. The data were collected for 3–5 areas of each specimen. The digitized roughness profiles were measured in two mutually perpendicular directions ($\parallel$) and ($\perp$).

Fig. 3 demonstrates a typical AFM relief of a LiF dissolution surface and "microroughness" profile for one of the 1000 digitized scans. Two different averaging techniques were used in the FNS analysis. In the first case, parameters $<h>$, $\sigma$, $L_{01}$, $L_1$, $n$, $2H_1$, $S_c(0)$, and $S_R(0)$ were calculated for each of the 100 randomly-picked scans, and then their values were averaged. In the second case, the profiles of the 100 randomly-picked scans were first averaged for each of the directions ($\parallel$) and ($\perp$), and then the parameters were calculated. The difference in the results produced by both techniques did not exceed 5%.

The calculated set of FNS parameters that characterize the AFM images of LIF surfaces for each of the directions ($\parallel$) and ($\perp$) are presented in Table 1. The analysis shows that, in contrast to the traditional statistics $<h>$ and $\sigma$, all FNS parameters are sensitive to variations in surface structure which depends on the features of chip "preparation" (hydrogenation, dissolution mode, anisotropy caused by the cleaving of a single crystal, etc.) [29,30].

*6.2.2. FNS parameterization of complex AFM images for surfaces with dendritic structures*

FNS parameterization is much more complex when various "structures" are formed on the surface. Consider such an example where FNS is used to parameterize AFM images of two dendritic (treelike) structures formed on mica surface from a solution of pluronic F-127 in chloroform (Fig. 4) and water (Fig. 5) [27]. A *Smena* scanning probe microscope manufactured by company *NT-MDT* (Zelenograd, Russia) was used. The images of surface fragments of film specimens were taken in the tapping mode using silicon cantilevers of *NSG* 11 series (*B* type, resonant frequency of 150 kHz, rounding radius of 10 nm). The number of scans along *x* (horizontal axis) and *y* (vertical axis) was 256.

Pluronics are surfactant compounds that represent block copolymers of ethylene oxide (A) and propylene oxide (B) of ABA type. For the pluronic F-127 manufactured by BASF (USA) with a molecular weight of 12600, the ratio of the number of ethylene oxide chains to the number of propylene oxide chains was 3.1:1. In chloroform solutions, pluronics do not form micelles and are kept as separate molecules (unimers) at any concentration. In aqueous solutions, pluronics form micelles if their concentration exceeds the critical concentration of micelle formation (~$5.5 \cdot 10^{-4}$ mole/l per macromolecule [45]). The internal (hydrophobic) nucleus of a pluronic micelle is made up of polypropylene oxide fragments while its external (hydrophilic) nucleus is formed by polyethylene oxide fragments [46]. Precipitates of pluronic F-127 were obtained from F-127 solutions with a concentration of $3 \cdot 10^{-6}$ M. The solutions were spread on the mica surface and then vaporized at room temperature, for ~1 minute for chloroform or 10-15 minutes for water.

Although the "dendritic" contours on AFM images of pluronic F-127 precipitates presented in Figs. 4 and 5 are much alike, the local structures of the precipitates are different. It is obvious that the larger fragmentariness of the structure formed in aqueous solution is associated with the pluronic micelle formation in water solution that occurs closer to the end of the process due to the increase in pluronic concentration. The fractal dimension of the presented images, which was determined using the "Box dimension" software program [47], which analyzes the dependence of the number of squares required to cover the figure on the size of the square sides, was found to be 1.78 for Fig. 3 and 1.70 for Fig. 4. As was pointed out above, the characterization of such complex structures by one parameter does not allow us to describe many of the features in their structural organization.

FNS parameterization of the AFM images was carried out using the second method of averaging described in section 6.2.1 [27]. FNS parameters were calculated for the averaged scans of image projections on coordinate axes $x$ and $y$. Projections of the images displayed in Figs. 4 and 5 on axes $x$ and $y$, which may be considered as "fingerprint profiles", are shown in Figs. 6 and 7. The corresponding power spectra $S(k)$ (for $L^{-1} < k \leq k_{max} = 64\,L^{-1}$) and difference moments $\Phi^{(2)}(\Delta)$ (for $0 \leq \Delta \leq \Delta_{max} = ¼\,L$) are demonstrated in Figs. 8-11. The argument intervals were selected from the following considerations. When the intervals are large, the individuality of the signal becomes more prominent. At the same time, the error in calculating $S(k)$ and $\Phi^{(2)}(\Delta)$ gets higher. So we selected some "optimum" values for which the error is not significant, yet the individual features are already discernible.

Each of the functions given in Figs. 8-11 is quite specific and can be regarded as one of the "patterns" that characterize the images in Figs. 4 and 5. These patterns are formed both by the resonant and chaotic components of the "fingerprint" profiles shown in Figs. 6 and 7. Hence, the FNS analysis carried out with Eqs. (11), (15), (26), and (27) introduced the "resonant" (positions $k_{0i}^{jk}$ of main "resonances", their "half-widths" $\gamma_i^{jk}$, and "intensities" $\alpha_i^{jk}$) as well as "chaotic" (correlation length, Hurst constant) parameters for the AFM images. The upper indices $j$ ($j = 1, 2$) and $k$ ($k = x, y$) point to the number of the image (1 for Fig. 4, 2 for Fig. 5) and coordinate axis ($x$ or $y$) on which the image profile is projected, respectively.

The values of resonance parameters $k_{0i}^{jk}$, $\gamma_i^{jk}$, and $\alpha_i^{jk}$ for the projections of images in Figs. 6 and 7, which were determined using Eq. (26) from functions $S(k)$ (Figs. 8 and 9), are listed in Tables 2 and 3. Only the largest (in $A_i^{jk}/\gamma_i^2$ magnitude) 8-9 resonances were considered. Parameters $\sigma^{jk}$, $H_1^{jk}$, and $L_1^{jk}$, which characterize the chaotic component of the projections of images in Figs. 6 and 7, were determined by fitting the functions $\Phi^{(2)}(\Delta)$ calculated using Eqs. (11), (15), and (27) and data in Tables 2 and 3 to the corresponding functions shown in Figs. 10 and 11 (solid lines). The resulting calculated functions are presented in Figs. 10 and 11 (dashed line), and the appropriate parameters $\sigma^{jk}$, $H_1^{jk}$, and $L_1^{jk}$ are listed in Table 4. It is seen that the functions calculated using the simple model expressions (15), (26), and (27) for a large enough number of resonance parameters, determined from independent power spectra $S(k)$, repeat the main features of the "original" functions $\Phi^{(2)}(\Delta)$.

As the calculated functions are sensitive to the values of the determined parameters, all parameters presented in Tables 2-4 can be considered as the "passport" characteristics of the images in Figs. 4 and 5. A comparative analysis of the data presented in Tables 2-4 shows that the determined parameters reflect the physical differences in the conditions of pluronic precipitate formation from aqueous and chloroform solutions. The higher values of correlation lengths $L_1^{2k}$ as compared to $L_1^{1k}$ may be attributed to the features of aggregation dynamics. In water solutions, hydrated pluronic micelles (larger aggregates) are formed at the final phases of precipitate crystallization. In chloroform solutions, the micelles are not formed at all.

The differences in the values of $H_1^{2k}$ and $H_1^{1k}$, which characterize the correlation of image fragments at relatively small distances ("short-range memory"), are especially illustrative. When pluronic precipitates are formed from water, their interactions, effected through hydrated micelle shells, disappear at the final stages of precipitate structure formation. Thus, the individual aggregates of pluronic molecules with characteristic dimensions of ~0.10 $\mu$m do not interact with each other despite the general relations between macroscopic precipitate fragments at correlation lengths $L_1^{2k} \sim 10$ $\mu$m. This "loss" of local links manifests itself in higher values of Hurst constant $H_1^{2x} = H_1^{2y} = 0.4$ for the case of water solution. For the case of chloroform solution, the constants are significantly lower: $H_1^{1x} = 0.15$, $H_1^{1y} = 0.3$. This difference in the values is attributed to the larger geometric asymmetry of the image in Fig. 4 as compared to the image in Fig. 5. This asymmetry also affects the values of correlation lengths $L_1^{1x}$ and $L_1^{1y}$, $L_1^{2x}$ and $L_1^{2y}$, though to a lesser degree.

FNS parameterization can be applied to the analysis of surface structures of almost any type because the number of introduced parameters can be easily increased by studying the difference moments of higher orders. Furthermore, since the coordinate axes may be chosen rather arbitrarily, the number of projection axes (and, subsequently, projections themselves) can be increased, which leads to an increase in the number of introduced parameters. Of course, in practice one needs to determine the "minimum" number of parameters that is adequate for solving a particular problem. FNS may be used to carry out this type of task as well.

*6.2.3. Other examples of FNS parameterization*

In the same manner, FNS may be used in the analysis of time series $V(t)$ to determine the parameters that characterize the dynamic state of complex systems. When the dynamics is most complex; that is, there are 10-15 specific resonances that play role in the evolution, the problems may be solved by analyzing functions $\Phi^{(2)}(\tau)$ and $S(f)$, which represent the "passport" patterns of evolution, and their changes in response to various factors.

For example, the high specificity of $\Phi^{(2)}(\tau)$ and $S(f)$ patterns obtained in the analysis of complex system states may be used to diagnose diseases at their initial stages. Such FNS application is demonstrated by Timashev et al. [34], who analyzed the electroencephalograms of two patients, a healthy ("normal") child and one with schizophrenic symptoms, using a frequency power spectrum $S_F(f)$ measured at C4 points and other FNS functions. FNS was also used to analyze the effects of different types of medical treatment on the dynamics of index finger tremor for a Parkinsonian patient [35]. The tremor change rate data were obtained with a special laser device. The signals were measured after two types of treatment: medication (L-Dopa) and electromagnetic stimulation (DBS) with the excitation of the electrode implanted in the brain of the patient. Flags "ON" and "OFF" were introduced to indicate the use or exclusion of electromagnetic stimulation or medication in the experiment. Four types of process conditions were studied: "ON-ON", both medication and stimulation are used; "ON-OFF", medication is used, stimulation is excluded; "OFF-ON", medication is excluded, stimulation is used; "OFF-OFF", both medication and stimulation are excluded. It was shown that the power spectra and difference moments calculated using the measured tremor signals for the Parkinsonian patient can be considered as informative patterns that allow one to evaluate the condition of the patient as well as determine the efficiency of medications.

*6.3. Dynamics of nonstationary processes*

*6.3.1. Precursors of electric breakdown in thin porous silicon films*

Consider the electric breakdown in semiconductor systems as an example of a catastrophic event. It is generally believed that electric breakdowns are preceded by some irreversible structural reconfigurations. Our goal is to locate the "forerunners", i.e. precursors, of such breakdowns, which could tell us in advance that some "dangerous" event is building up. It is assumed that there is a correlation between the structural changes preceding the breakdown and variations in electric current and voltage.

Fluctuations of electric current in porous silicon at a fixed anodic voltage were studied by Parkhutik et al. [32]. The operating temperature was linearly increased with time from 20 to 250°C, which corresponds to the breakdown conditions. A dye was applied to the surface of the porous silicon specimen to visualize possible structural microscopic changes on the surface that are caused by the electric breakdown. After applying the dye, thin gold "spots" (~2 mm$^2$) were deposited on the surface to provide an electric contact with a corresponding electrode.

Micrographs of the specimen surface before and after the electric breakdown are illustrated in Fig. 12. A typical curve for the dependence of the electric current density on time in the conditions of linearly increasing temperature and applied voltage of $U = 70$ V is depicted in Fig. 13a. The variation of nonstationary indicator $C_F$ with time is shown in Fig. 13b. The indicator was evaluated with Eq. (28) for the high-frequency part of the electric current density at $\alpha = 0.5$ using "averaging windows" of $T = 70$ s and $\Delta T = 20$ s.

It is seen from Fig. 13b that the electric breakdown is preceded by a peak value (spike) of the nonstationarity criterion (marked with arrow *1*), which arises 20–25 s before the breakdown. The peak may reflect some structural changes in the specimen of porous silicon that take place shortly before the electric breakdown. In addition to this singularity, there is a considerable peak in function $C_F$ (marked with arrow *2* in Fig. 13b), which may be attributed to some earlier structural reconfigurations. It can be suggested that the "catastrophic" changes in the surface structure of the porous-silicon film, which are shown in Fig. 12b, started building up some time before the breakdown. In this example, this time period is about 200 s.

Hence, the structural changes preceding the breakdown event were successfully captured using the FNS indicators of nonstationarity. Consider another example to find out if this suggestion is valid for other "catastrophic" phenomena.

*6.3.2. Analysis of geoelectrical signals measured in seismic areas*

The anomalous variations in some geophysical parameters, such as local fluid flows (water, saline components, gases), medium electrochemical variables, geoacoustic and geoelectromagnetic streams, geophysical ground fields, are believed to be related to system reconfiguration that precedes earthquakes. For example, some geophysical parameters measured in seismic areas varied anomalously shortly before the earthquakes took place, which was attributed to stress and strain changes [48,49]. Thus, a lot of recent studies on locating intermediate and short term seismic precursors have focused on the analysis of time series for geophysical parameters.

In addition, geoelectrical parameters measured in seismic areas can also help understand various seemingly complex phenomena related to seismic activity [50,51]. For example, variations in the stress and fluid flow fields can produce changes in the geoelectrical field, resistivity, and other electrical parameters [52]. Therefore, the analysis of induced fluctuations may provide information on the governing mechanisms both in normal conditions and during intense seismic activity.

Presently, the use of electrical precursors in earthquake prediction is still to a large extent empirical because of the difficulties in understanding the physical mechanisms of geophysical precursory phenomena. In order to assess the predicting power of geoelectrical signals as indicators of earthquake preparation [53,54], it is necessary to know whether these parameters are related to dynamic characteristics of active tectonics and if there is a significant correlation between seismic events and geoelectrical temporal fluctuations. Obviously, the existence of such a correlation can be established only after the dynamic characterization of the geoelectrical signals is accomplished. A short review of the existing methods of analyzing the time series for geoelectrical signals is presented by Telesca et al. [55].

The time series of hourly self-potential measurements recorded in 2002 at station Giuliano located in a seismically active area in southern Italy were analyzed using FNS [55]. Technically, a geoelectrical or self-potential time series is a sequence of voltage differences measured at a selected sampling interval using a receiving electrode array. During the geoelectrical soundings, when the electric current is introduced into the ground, the self-potential variations represent the noise. On the other hand, the main signal is measured using a passive measurement technique (i.e., without an energizing system). To avoid self-polarization effects, ceramic electrodes, which are ceramic vessels filled with a saturated solution of copper sulphate, were used.

Hourly time variations in geoelectrical signal $V(t)$ measured at station Giuliano during January-September 2002 are illustrated in Fig. 14. The low frequency component $V_R$ and high frequency component $V_F$ obtained using the diffusion equation technique described in Chapter 4 are depicted in Fig. 15. The criteria $C_F$ calculated for two averaging intervals $T=720$ hours and $T=1080$ hours with $\Delta T=24$ hours is shown in Fig. 16. The figure demonstrates a strong variability of criterion $C_F$ with several sharp changes. This implies that the signal is highly nonstationary. Furthermore, the negative spikes for $T=720$ and $T=1080$ are shifted with respect to each other. Therefore, they are related to the duration of the averaging interval $T$ and are "side effects" of FNS. At the same time, the time position of the positive spikes is invariant with respect to the averaging interval $T$ and depends only on signal characteristics. Thus, the second type of the spikes may be considered as informative.

In order to find correlations with earthquakes, one needs to select only those seismic events that could be responsible for significant changes in the time dynamics of the geoelectrical signal. Thus, only the earthquakes that could be responsible for strain effects in the area under study were chosen. The stress field produces cracks in the rock volumes, which triggers changes in fluid pressure. This leads to an underground charge motion, and the latter results in anomalies in the electrical field on the surface, which were observed only when the preparation region was near the measuring station. It is necessary to discriminate the major events (i.e., earthquakes responsible for significant geophysical variations in the rock volume of the area under study) from all other seismic sequences that occurred in the area surrounding the measuring station. To this regard, an empirical formula introduced by Dobrovol'skiy [56,57] was used:

$r = 10^{43M}$,

where $M$ is the magnitude and $r$ is the radius, in kilometers, of the area in which the effects of the earthquake are detectable. The earthquakes with $r$ greater than the distance between the epicenter and the measuring station were selected.

Fig. 17 shows the criteria $C_F$ along with the magnitudes of seismic events in this area, taken from the INGV catalogue, that meet the Dobrovol'skiy's rule. Also, the mean and $\sigma$ range of the criteria are plotted. Apparently, there is a correlation between the positive spikes in the criterion $C_F$ and the earthquakes. Specifically, some positive sharp changes (spikes) in the criterion seem to precede the occurrence of a seismic event selected using the Dobrovol'skiy's rule. Furthermore, no strong variations in $C_F$ were detected during the largest interevent time period between the fourth and fifth earthquakes, which further supports the suggestion that positive spikes in $C_F$ may be used as precursors of earthquakes.

The same conclusion was obtained by Descherevsky et al. [58], who used FNS to analyze the time series of electrochemical potential measured in the period from 1979 to 1991 at two different measuring locations in Tajikistan.

*6.4. Determination of correlation links in the dynamics of distributed systems*

*6.4.1. Dynamics of membrane-potential "interelectrode" correlations in electromembrane systems with overlimiting current density*

Consider the application of FNS to the analysis of electric potential fluctuations in an electromembrane system [28,59], which is an illustrative example of finding correlation links in the dynamics of distributed systems. A four-chambered unstirred glass cell was used to record the electric noise [60]. A schematic of the measuring unit is shown in Fig. 18. Each of the chambers represented a glass cylinder with a length of 25 mm and internal diameter of 15 mm. The chambers with measuring electrodes were separated from the anode and cathode chambers by ultrafiltration membranes to prevent the contamination of the system by the products of electrode reactions. Polarizing cylindrical electrodes made of graphite were placed in the external chambers. A cation-exchange aromatic polyamide membrane with a thickness of 55 $\mu$m was vertically positioned, and separated two internal chambers in the electrodialyzer.

The electric potential noise near the membrane surface was measured using a holder with microelectrodes, which limited the measurement area of the membrane surface to a rectangle with dimensions of 10x1 mm$^2$. The measuring electrodes made of a copper wire with a diameter of 100 μm were rigidly mounted on the longer sides of the rectangular holder at a distance of ~200 μm from the membrane surface. The silverized ends of the wire, on which a layer of AgCl was formed, were used as the working surfaces of the electrodes. The distance between the adjacent electrodes was in the range from 0.5 to 4 mm. Two sets of 8 measuring electrodes each were placed on the longer sides of the rectangular holder. An Ag/AgCl electrode positioned in the measuring chamber at a distance of 15 mm from the membrane was used as the reference electrode. The reference electrode along with all 16 measuring electrodes was placed on the same side of the membrane under study. A 0.1M NaCl solution was used as the electrolyte. All measurements were made at a constant temperature of 22°C.

The noise experiments were carried out in the galvanostatic mode. The polarizing electrodes were connected to a custom-made regulated DC source. The potential differences between the reference electrode and 16 measuring electrodes were synchronously recorded using a 12-bit multichannel L-305 ADC board (L-Card Co., Moscow, Russia) with an input resistance of 1 Mohm. Fluctuations in the measured voltages were observed only when the measuring electrodes were positioned on the membrane side facing the positive polarizing electrode. No fluctuations were observed in the opposite case, when the measuring electrodes were positioned on the membrane side facing the cathode electrode. The data listed below were obtained for the case in which the longer side of the membrane was positioned horizontally (Fig. 18). The microelectrodes shown in this figure were numbered sequentially starting with the top row from left to right (1, 2, 3, etc.), and (9, 10, 11, etc.) for the bottom row. With such numbering scheme, electrode 9 faces electrode 1, 10 faces 2, and so on. The cross-correlators $q_{ij}(\tau, \theta_{ij})$ for voltage time series calculated for electrodes $i$ and $j$ will be referred to as "cross-correlators $i$-$j$".

Fig. 19 shows an example of the chaotic dynamics for signals $V_2(t)$, $V_3(t)$, $V_4(t)$, and $V_{11}(t)$ measured for $T_{tot} = 4096 f_0^{-1} \approx 41$ s. The time series were derived from the measured values of the membrane potentials by removing the linear trend. The discretization frequency was 100 Hz. The total number of measured values for each voltage time series was 4096.

Figs. 20-21 present two-parametric, $\tau$- and $\theta_{ij}$ -dependences of cross-correlators 2-3, 4-3, 3-11, 1-3, and 2-4 given by Eq. (29). Each of the functions $q_{ij}(\tau, \theta_{ij})$ was calculated for time intervals $T = 500 f_0^{-1}$, where $f_0$ = 100 Hz is the discretization frequency. The data in Fig. 20 are given for intervals I (1-500) $f_0^{-1}$, II (1201-1700) $f_0^{-1}$, III (2601-3100) $f_0^{-1}$, IV (3001-3500) $f_0^{-1}$. It is reasonable to suggest that the measured fluctuations in the local values of electric potentials are related to the fluctuations in hydrodynamic flows near the microelectrodes. The measured functions reflect a complex dynamics of temporal variations of hydrodynamic flows in the electrolyte solution near the membrane surface. In this case, the locations of maxima in the dependences of $q_{ij}(\tau, \theta_{ij})$ on $\theta_{ij}$ indicate what direction the hydrodynamic flows are taking. If the maximum of $q_{ij}(\tau, \theta_{ij})$ takes place at $\theta_{ij} = 0$, this implies that either the electrolyte flow going from the $i$-th electrode to the $j$-th electrode counterbalances the opposite, from $j$-th to $i$-th, flow, or the "external" flows from other near-membrane areas are producing the same hydrodynamic conditions in the proximities of the $i$-th and $j$-th electrodes. In the latter case, the characteristic size $l$ of local hydrodynamic pulsations should be comparable with the distance $l_{ij}$ between the $i$-th and $j$-th electrodes; that is, $l \sim l_{ij}$. When $l \gg l_{ij}$, $q_{ij}(\tau, \theta_{ij})$ should be independent of $\theta_{ij}$. In this case, the signals $V_i(t)$ and $V_j(t)$ do not affect each other: their fluctuation dynamics is determined by "side" electrolyte flows.

The data in Fig. 20a demonstrate that the signals $V_2(t)$ and $V_3(t)$ change in antiphase ($q_{23} < 0$) on time interval I, and do not affect each other. The situation changes on time interval II (Fig. 20b): the electrolyte flows from the 2nd to the 3rd electrode bring about a symbatic (in average) variation of $V_3(t)$ with $V_2(t)$; that is, one can see a large positive correlation for $\theta_{23} > 0$, a small correlation for $\theta_{23} \approx 0$, and a large negative correlation for $\theta_{23} < 0$. At the same time, the flows from the 3rd to the 2nd electrode produce an antibatic variation of $V_3(t)$ with $V_2(t)$, which is complementary to the cross-correlation 2-3. On interval III (Fig. 20c), the observed maximum in the dependence of $q_{23}(\tau, \theta_{23})$ on $\theta_{23}$ may be attributed to the propagation of local pulsations $l \sim l_{23} = 0.5$ mm rather than the mutual compensation of electrolyte flows between electrodes 2 and 3. This conclusion is supported by the fact that there are correlations between signals $V_1(t)$ and $V_3(t)$; i.e., electrodes 1 and 3, on that interval (Fig. 21a), while there are none between electrodes 2 and 4 (Fig. 21b). There are no correlations between signals $V_2(t)$ and $V_3(t)$ on interval IV.

The data in Fig. 20e show that the "external" flows from other near-membrane areas are producing the same hydrodynamic conditions in the proximities of electrodes 4 and 3 on interval I. This interpretation of the maximum at $\theta_{43} = 0$ is based on the observation that the variations of cross-correlators $q_{43}(\tau, \theta_{43})$ and $q_{311}(\tau, \theta_{311})$ on this interval are nearly equal to each other, which implies the presence of a "fluctuation field" acting uniformly on electrodes 3, 4, and 11. On interval IV, the flows from electrode 4 toward 3 exceed the reverse flows. There are almost no correlations between signals $V_4(t)$ and $V_3(t)$ on interval III. The data in Figs. 20k-p demonstrate that the opposite flows between electrodes 3 and 11 are equal to each other throughout the whole time interval $T_{tot}$. The above figures, particularly Fig. 20e and Fig. 20g, imply that the maximum velocity of hydrodynamic flows between the adjacent electrodes, which are 0.05 cm apart from each other, is ~ 0.2 cm/s.

The complete analysis of correlations between the signals of all 16 electrodes makes it possible to determine the features of temporal electrolyte-flow redistribution in the near-membrane area of the electromembrane system.

*6.4.2. Other examples of correlation link analysis in the dynamics of distributed systems*

The two-parametric cross-correlators given by Eq. (29) may also be used in the analysis of various medical characteristics, such as electroencephalograms (EEG). The cross-correlators for EEG signals measured at C4 and O2 points for two patients, a healthy ("normal") child and one with schizophrenic symptoms, are presented by Timashev et al. [34]. The dramatic differences observed in the behavior of the corresponding cross-correlators indicate that the FNS approach may be considered as a new tool for early diagnostics of various brain diseases, such as schizophrenia, Alzheimer, Parkinson, and many others.

**7. Conclusions**

The use of autocorrelator (1) as a basis for extracting phenomenological information about the evolution dynamics and structure of complex systems is virtually a new epistemological paradox; i.e., a situation when we can apply some basic concepts without understanding them. One of the most illustrative examples of such paradox was the concept of probability [61]. "Its addition to cognition led to radical transformations in the scientific picture of the world, style of scientific thinking, and basic models of world creation and their perception" [62]. Some scientists even discussed the idea of "probabilistic revolution" in science [63].

The level of information contained in autocorrelation function $\psi(\tau)$, which is associated with signal correlation links on different spatiotemporal intervals, is more profound than the one in probability density functions. However, the correlation functions $\psi(\tau)$ and their generalizations to higher orders are traditionally used only as subroutines in general theoretical schemes of statistical physics. Functions $\psi(\tau)$ are rarely used as the basis of signal analysis. FNS allows us to overcome the difficulties associated with direct extraction of information from autocorrelator $\psi(\tau)$ by using its special "projections", Fourier transform and structural functions. The FNS approach makes it possible to determine both system-specific (resonant) and non-specific (chaotic) features of signals even if the principles of the system generating the signals are not well understood (epistemological paradox). In fact, this general knowledge about the dynamics of a system and its structure that is acquired with FNS may provide the basis for further understanding of the system.

The presented methodology for analyzing the state and dynamics of complex systems may be applied to studies of various physical and physicochemical problems (dynamics of phase transitions in extended systems, kinetics of heterogeneous catalytic and electrocatalytic transformations, and the like), monitoring of natural objects, prediction of climatic and ecosystematic changes, chemical engineering (scaling of turbulent flows in fluidized-bed units or membrane systems), parameterization of surface structures (metrology on nano- and microscales), astrophysics (analysis of variations in the activity of stellar and quasi-stellar objects), geophysics (prognostication of atmospheric events, seismic activity, dynamics of processes in Earth's magnetosphere), medicine (early diagnosis of various diseases, for example, unsolved problems with early diagnosis of Alzheimer and Parkinson diseases as well as schizophrenia based on electroencephalogram data), and genetics (analysis of dynamic mutations, and possibly finding traces of hidden information in non-coding genome sequences). It should be noted that FNS can succeed even when no complete set of data about the system under study is available. As this is true for many natural systems, the methodology can become a helpful analytical tool in solving practical problems.

The FNS methodology may be considered as the basis for a new kind of information technology (IT), Informative Noise Spectroscopy (INS). The traditional IT deals with the use of electronic computers and computer software to convert, store, protect, process, transmit, and retrieve information. It is generally based on Shannon's presentation of information in terms of bits. One of the exceptions is the emerging field of quantum computing with block-matrix "qubits" as units of information. In FNS, the information is analyzed in terms of "block structures" related to irregularities of dynamic variables, such as spikes, jumps, discontinuities in derivatives on every spatiotemporal level of the system hierarchy. This opens up new opportunities in the extraction of information from complex signals. It is the interpretation of the irregularities as information carriers that allows FNS to classify all information contained in chaotic series, as well as reliably extract desired parts of information.

**Acknowledgement**

This study was supported in part by the Russian Foundation for Basic Research, project no. 05-02-17079.

**Tables**

Table 1. Structural and dynamic parameters of LiF dissolution surface

| $T$,°C | Scan direction | $h$, μm | $\sigma$, μm | $S_{cR}(0)$, nm² μm | $S_c(0)$, nm² μm | $H_1$ | $2H_1+1$ | $n$ | $L_{01}$, μm | $L_1$, μm |
|---|---|---|---|---|---|---|---|---|---|---|
| 30 | ⊥ | 0.33 | 0.08 | 96.5 | 299 | 0.93 | 2.86 | 3.62 | 1.14 | 0.99 |
| 30 | ∥ | 0.32 | 0.10 | 171 | 385 | 0.68 | 2.34 | 2.61 | 1.28 | 2.43 |
| 40 | ⊥ | 0.31 | 0.10 | 200 | 432 | 0.94 | 2.88 | 3.73 | 1.18 | 0.86 |
| 40 | ∥ | 0.31 | 0.11 | 394 | 640 | 0.60 | 2.16 | 2.38 | 1.33 | 2.06 |
| 50 | ⊥ | 0.34 | 0.11 | 356 | 740 | 0.91 | 2.84 | 3.69 | 1.07 | 0.81 |
| 50 | ∥ | 0.34 | 0.11 | 567 | 861 | 0.47 | 1.94 | 1.92 | 1.38 | 2.91 |

Table 2. The values of resonance parameters for the spatial series in Fig. 4

| $I$ | *$\alpha_i^{1x}$, $10^{-2}$nm² | $k_{0i}^{1x}$, μm$^{-1}$ | $\gamma_i^{1x}$, μm$^{-1}$ | **$\alpha_i^{1y}$, $10^{-2}$нм² | $k_{0i}^{1y}$, μm$^{-1}$ | $\gamma_i^{1y}$, μm$^{-1}$ |
|---|---|---|---|---|---|---|
| 1 | 0.229 | 1.58 | 0.157 | 1.343 | 2.22 | 0.672 |
| 2 | 1.175 | 2.53 | 0.16 | 1.076 | 4.12 | 0.314 |
| 3 | 0.473 | 3.16 | 0.157 | 0.5 | 4.75 | 0.314 |
| 4 | 2.29 | 4.43 | 0.314 | 0.188 | 6.34 | 0.151 |
| 5 | 1.053 | 5.7 | 0.317 | 0.911 | 7.6 | 0.628 |
| 6 | 0.519 | 6.66 | 0.157 | 0.258 | 8.28 | 0.157 |
| 7 | 0.458 | 7.6 | 0.157 | 0.31 | 8.86 | 0.314 |
| 8 | 0.977 | 8.3 | 0.314 | 0.163 | 10.4 | 0.157 |

* Values of $\alpha_i^{1x}$ are obtained by multiplying all parameters $A_i^{1x}$, determined from functions $S_{tot}^{1x}(k)$, by factor $q^{1x} = 0.6$ (see the text)
** Values of $\alpha_i^{1y}$ are obtained by multiplying all parameters $A_i^{1y}$, determined from functions $S_{tot}^{1y}(k)$, by factor $q^{1y} = 0.57$ (see the text).

Table 3. The values of resonance parameters for the spatial series in Fig. 5

| $I$ | $*\alpha_i^{2x}$, $10^{-2}$nm$^2$ | $k_{0i}^{2x}$, μm$^{-1}$ | $\gamma_i^{2x}$, μm$^{-1}$ | $**\alpha_i^{2y}$, $10^{-2}$nm$^2$ | $k_{0i}^{2y}$, μm$^{-1}$ | $\gamma_i^{2y}$, μm$^{-1}$ |
|---|---|---|---|---|---|---|
| 1 | 7.435 | 1.65 | 0.534 | 1.339 | 1.65 | 0.163 |
| 2 | 0.954 | 3.63 | 0.188 | 1.394 | 2.31 | 0.314 |
| 3 | 0.756 | 4.62 | 0.17 | 3.48 | 2.97 | 0.314 |
| 4 | 2.27 | 5.94 | 0.16 | 0.316 | 4.90 | 0.163 |
| 5 | 0.995 | 6.59 | 0.157 | 1.279 | 5.27 | 0.314 |
| 6 | 0.806 | 7.60 | 0.157 | 1.273 | 7.60 | 0.314 |
| 7 | 3.496 | 8.92 | 0.5 | 0.618 | 8.92 | 0.314 |
| 8 | 2.519 | 10.9 | 0.5 | 0.347 | 10.24 | 0.157 |
| 9 | 1.923 | 13.9 | 0.35 | – | – | – |

\* Values of $\alpha_i^{2x}$ are obtained by multiplying all parameters $A_i^{2x}$, determined from functions $S_{tot}^{2x}(k)$, by factor $q^{2x} = 0.24$ (see the text)
\*\* Values of $\alpha_i^{2y}$ are obtained by multiplying all parameters $A_i^{2y}$, determined from functions $S_{tot}^{2y}(k)$, by factor $q^{2y} = 0.57$ (see the text)

Table 4. Parameters $\sigma^{jk}$, $H_1^{jk}$, and $L_1^{jk}$

| $j; k$ | $\sigma^{jk}$, nm | $H_1^{jk}$ | $L_1^{jk}$, μm |
|---|---|---|---|
| 1; $x$ | 0.19 | 0.15 | 5.0 |
| 2; $x$ | 0.51 | 0.40 | 10.0 |
| 1; $y$ | 0.49 | 0.30 | 7.7 |
| 2; $y$ | 0.47 | 0.40 | 15.0 |

**Figure captions**

Fig. 1. Typical curves of functions (a) $\Phi_c^{(p)}(\tau)$ and (b) $S_c(f)$ for chaotic signals $V(t)$ without resonant components.

Fig. 2. Schematic of "random walk" evolution.

Fig. 3. (a) AFM image of a 50x50 $\mu m^2$ surface fragment of a lithium fluoride single crystal treated with hydrogen and subjected to surface dissolution in water at 25°C; (b) typical AFM relief (profile of roughnesses measured along the normal $z$ to the surface): 430-th of 1000 scans digitized along axis $x$.

Fig. 4. AFM image of the pluronic precipitate from chloroform solution on mica surface.

Fig. 5. AFM image of the pluronic precipitate from water solution on mica surface.

Fig. 6. Projections of the AFM image for pluronic precipitate in Fig. 4 on the $x$ (a) and $y$ (b) axes.

Fig. 7. Projections of the AFM image for pluronic precipitate in Fig. 5 on the $x$ (a) and $y$ (b) axes.

Fig. 8. Power spectra for projections of the AFM images of pluronic precipitate in Fig. 4 on the $x$ (a) and $y$ (b) axes.

Fig. 9. Power spectra for projections of the AFM images of pluronic precipitate in Fig. 5 on the $x$ (a) and $y$ (b) axes.

Fig. 10. Difference moments $\Phi^{(2)}(\Delta)$ for projections of the AFM images of pluronic precipitate in Fig. 4 on the $x$ (a) and $y$ (b) axes, which were calculated with Eqs. (3) and (10) for curve 1 and Eqs. (11), (15), and (27) using the values of parameters listed in Tables 2-4 for curve 2.

Fig. 11. Difference moments $\Phi^{(2)}(\Delta)$ for projections of the AFM images of pluronic precipitate in Fig. 5 on the $x$ (a) and $y$ (b) axes, which were calculated with Eqs. (3) and (10) for curve 1 and Eqs. (11), (15), and (27) using the values of parameters listed in Tables 2-4 for curve 2.

Fig. 12. Microphotographs of the surface of the dyed porous silicon specimen covered with a thin gold film: (a) before and (b) after electric breakdown.

Fig. 13. (a) Dependence of measured current density on time in the conditions of increasing temperature and applied voltage of $U = 70$ V up to the moment of electric breakdown; (b) time dependence of nonstationarity factor $C_F(t_{k+1})$ evaluated with Eq. (28) for the high-frequency part of the measured electric current density at $\alpha = 0.5$ using "averaging windows" of $T = 70$ s and $\Delta T = 20$ s.

Fig. 14. Geoelectrical signal variation recorded at station Giuliano in 2002.

Fig. 15. (a) Low-frequency $V_R$ and (b) high-frequency $V_F$ components of the signal in Fig. 14.

Fig. 16. Criterion $C_F$ for two values of averaging interval $T$, 720 and 1080 hrs, with $\Delta T$=24 hrs.

Fig. 17. Criterion $C_F$ along with magnitudes of the earthquakes that occurred in the area within the observation period, which were selected using the Dobrovol'skiy's rule: (a) $T$=720 hrs, (b) $T$=1080 hrs.

Fig. 18. Schematic of the unit for measuring potential fluctuations in an electromembrane system at different points: (1) computer, (2) multichannel analog-digital converter, (3) membrane under study, (4) measuring electrodes, (5) Ag/AgCl reference electrode, (6) ultrafiltration membranes, (7) polarizing electrodes, (8) DC source.

Fig. 19. Variations of signals $V_2(t)$, $V_3(t)$, $V_4(t)$, and $V_{11}(t)$ within time interval $T_{tot} = 4096 f_0^{-1} \approx 41$ s ($V_2^0 = 0\,\text{V}$, $V_3^0 = 0.5\,\text{V}$, $V_4^0 = 1.0\,\text{V}$, $V_{11}^0 = 1.5\,\text{V}$).

Fig. 20. Cross-correlators for electrodes (a, b, c, d) 2-3, (e, f, g, h) 4-3, and (k, l, m, p) 3-11: (a, e, k) on time interval I with $T = (1\text{-}500) f_0^{-1}$; (b, f, l), II with $T = (1201\text{-}1700) f_0^{-1}$; (c, g, m), III with $T = (2601\text{-}3100) f_0^{-1}$; (d, h, p), IV with $T = (3001\text{-}3500) f_0^{-1}$; $f_0 = 100$ Hz. Parameters $\tau$ and $\theta$ are given in units of $f_0^{-1}$.

Fig. 21. Cross-correlators for electrodes (a) 1-3 and (b) 2-4 on time interval III with $T = (2601\text{-}3100) f_0$; $f_0 = 100$ Hz. Parameters $\tau$ and $\theta$ are given in units of $f_0^{-1}$.

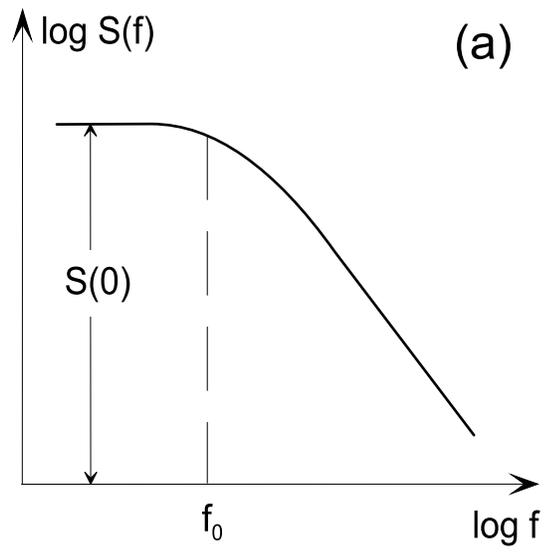
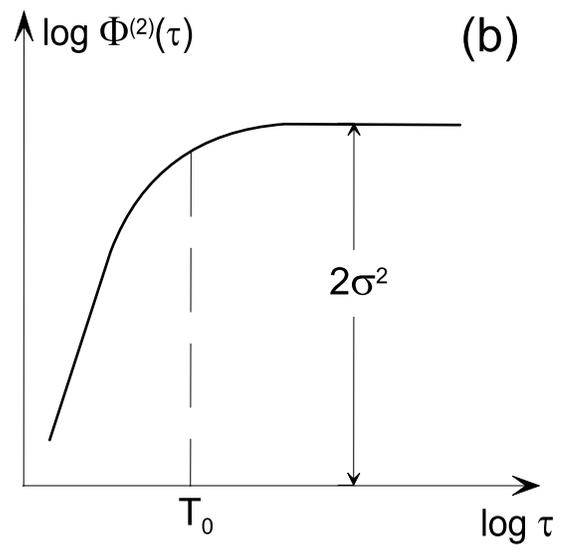

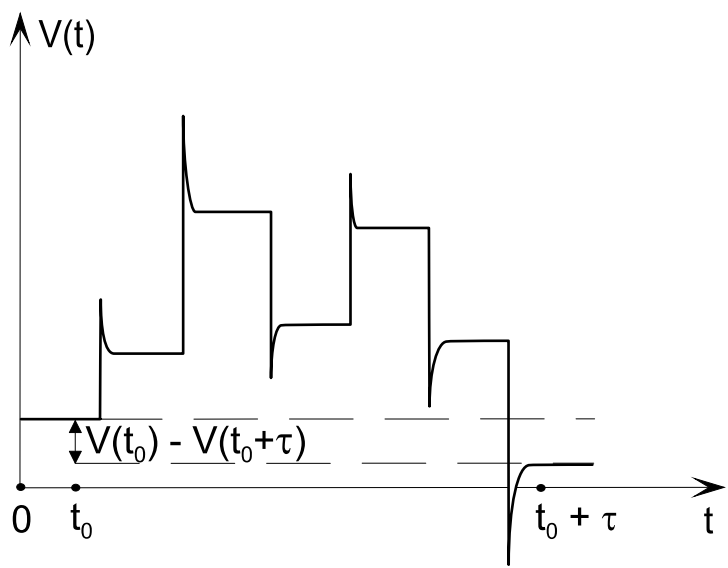

(a)
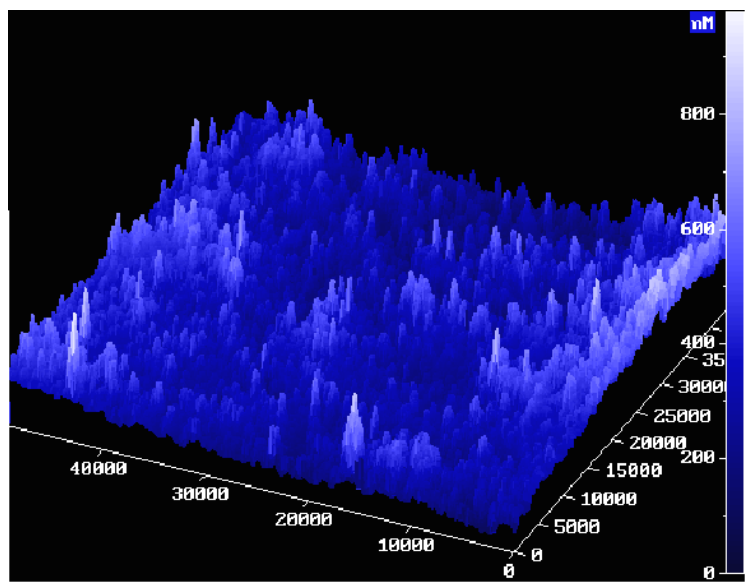

(b)
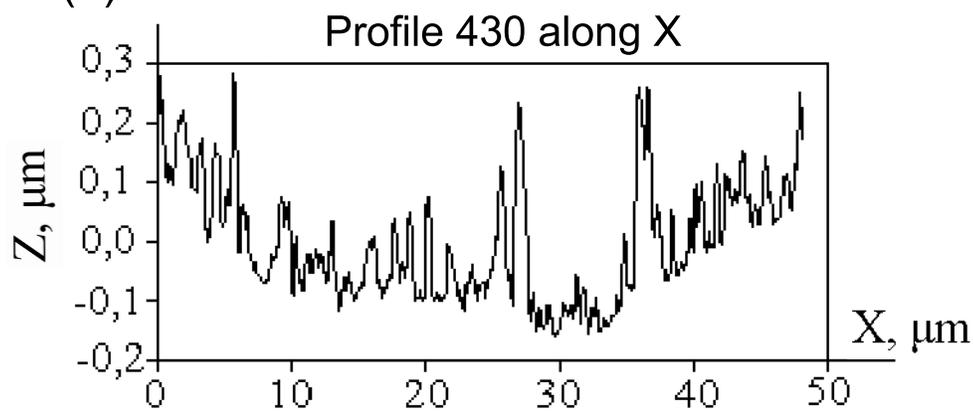

**Figure 4**

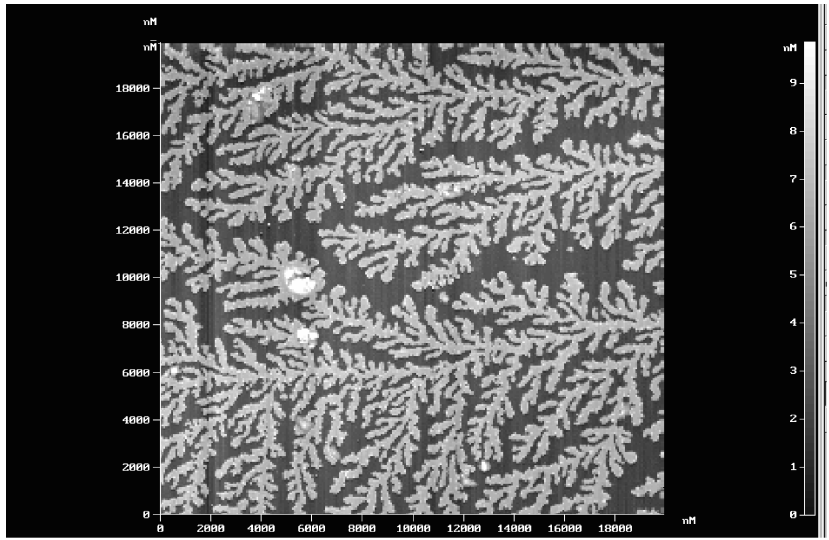

**Figure 5**

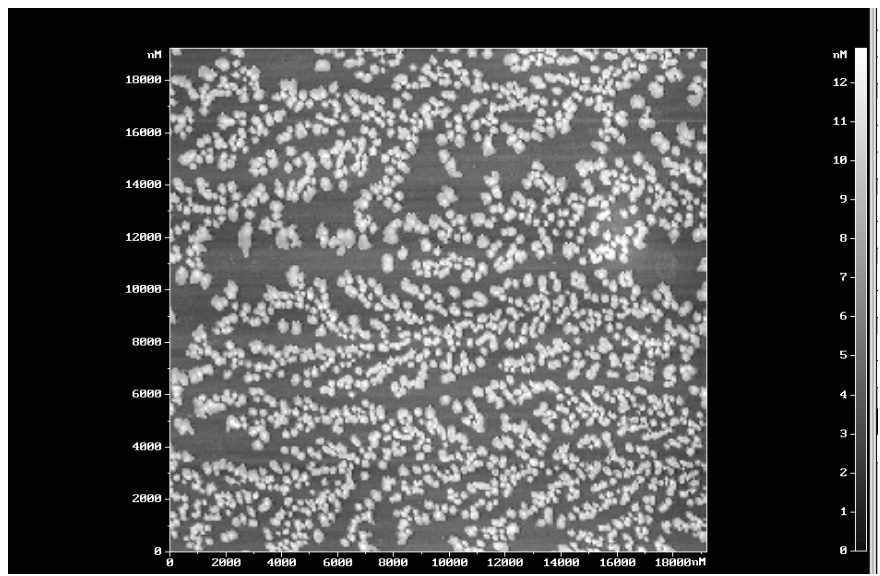

**Figure 6**

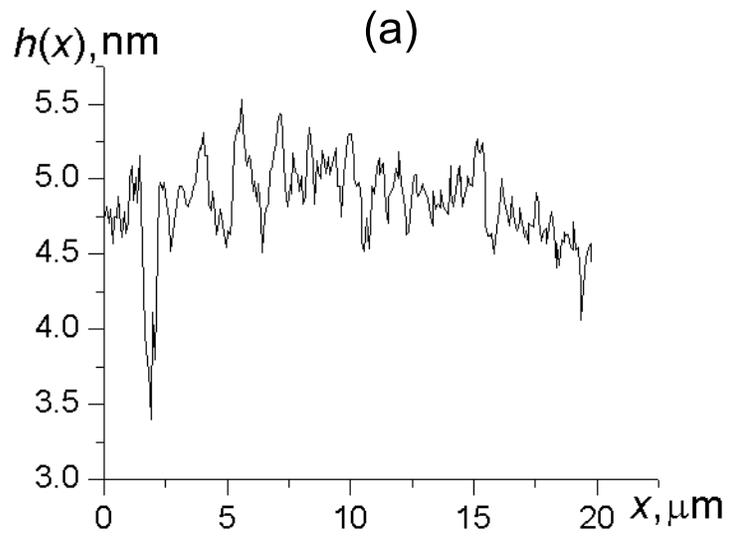 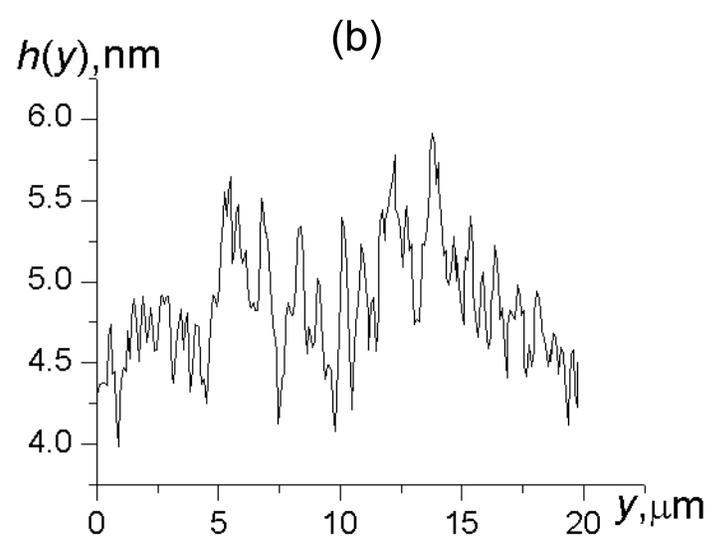

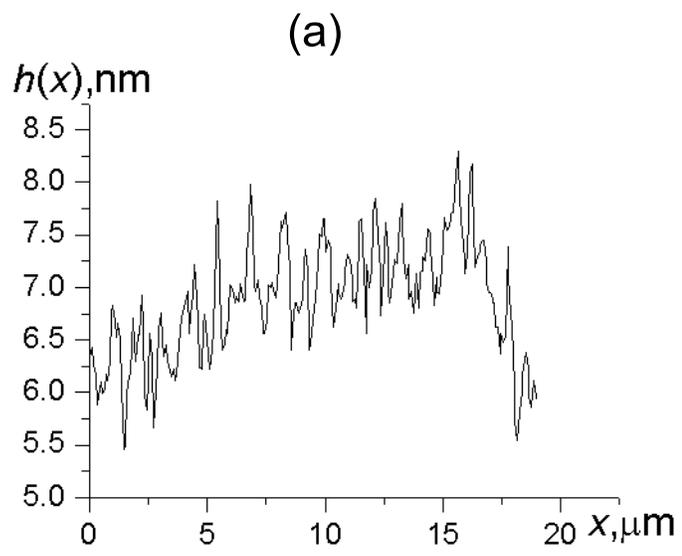 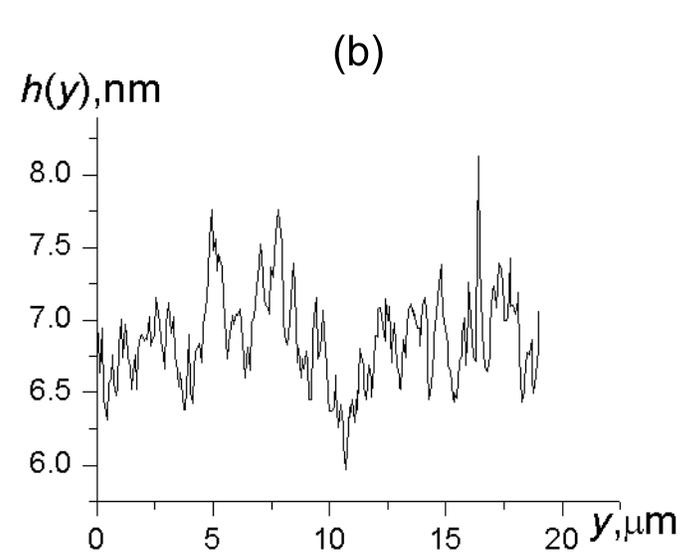

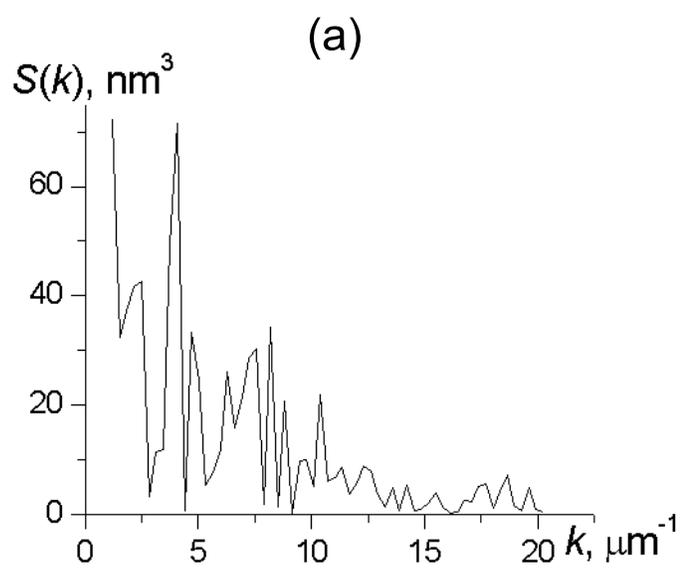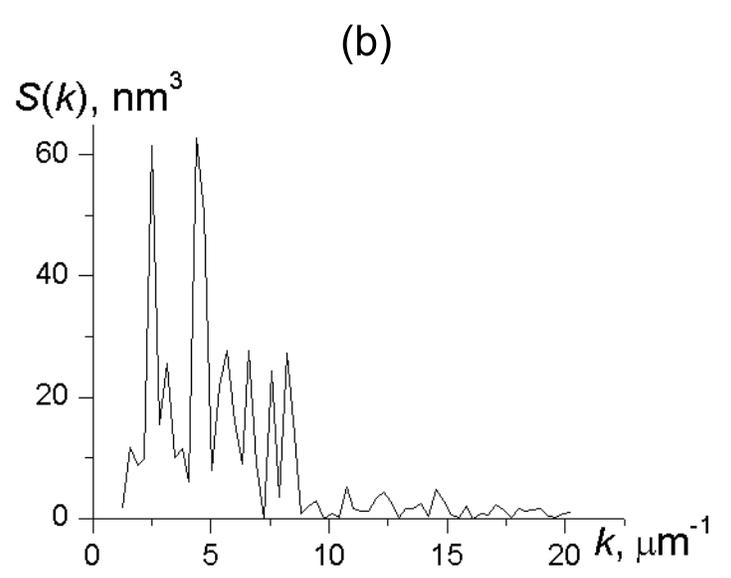

**Figure 9**

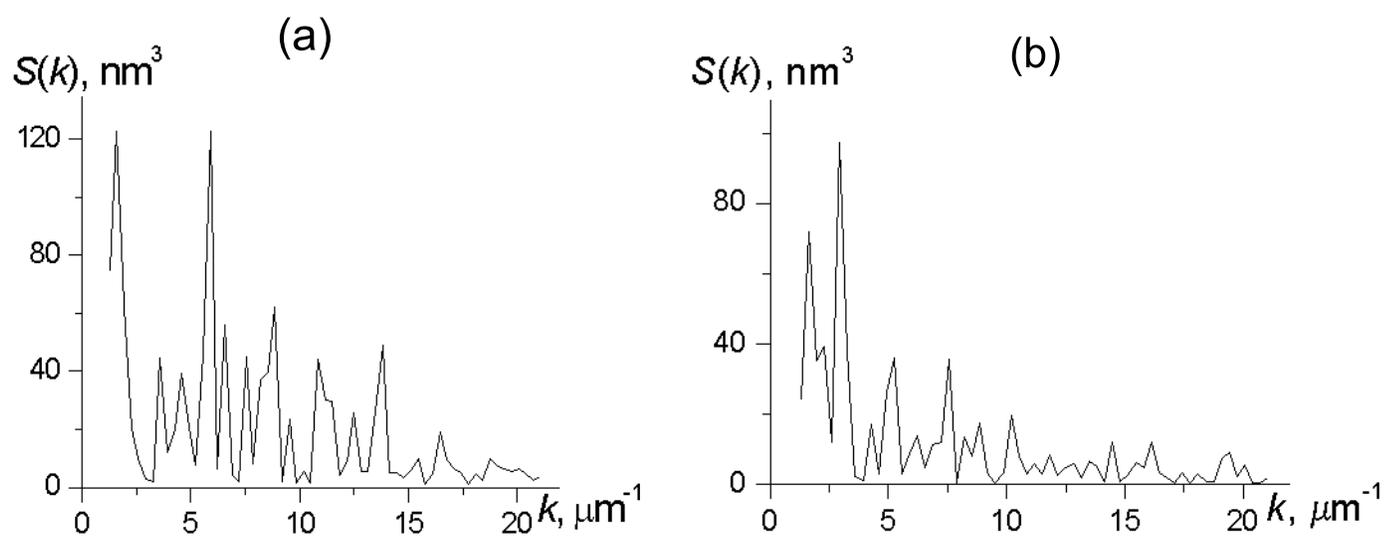

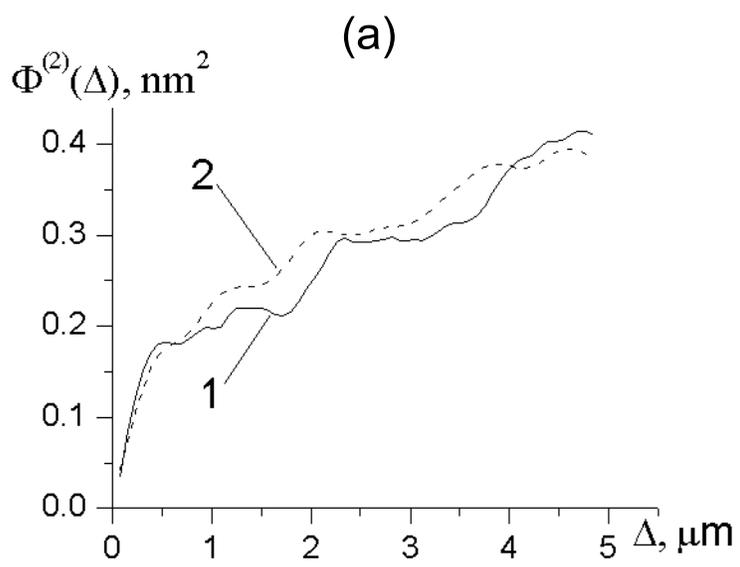 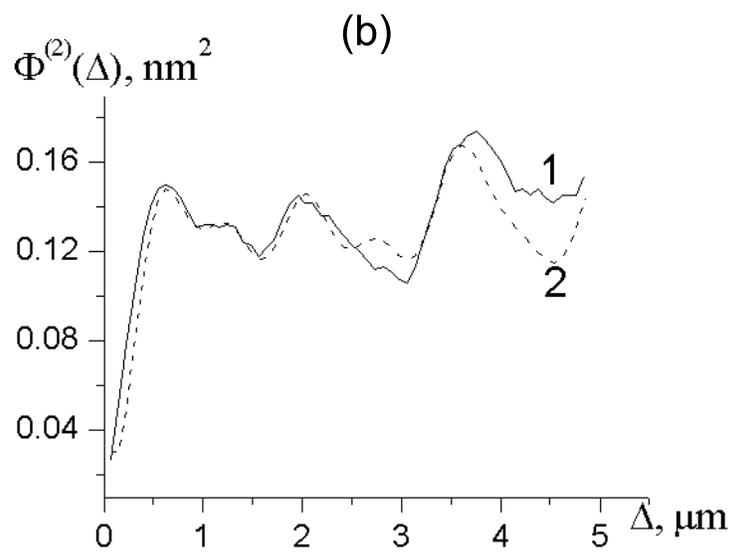

**Figure 11**

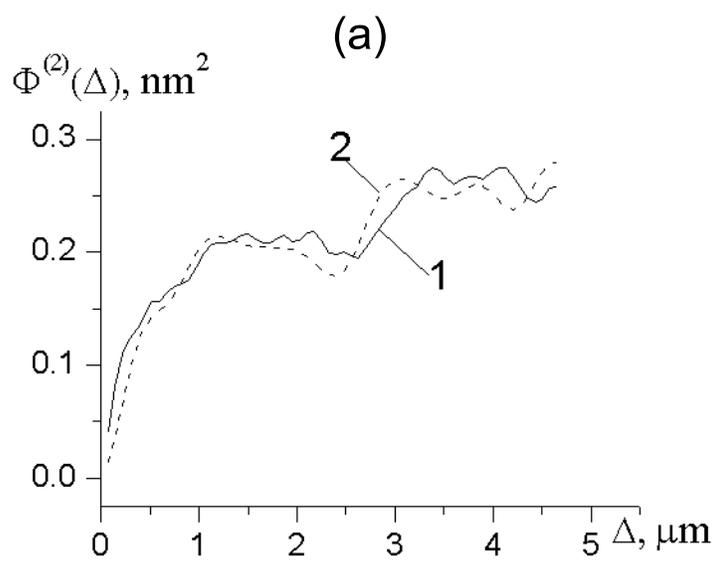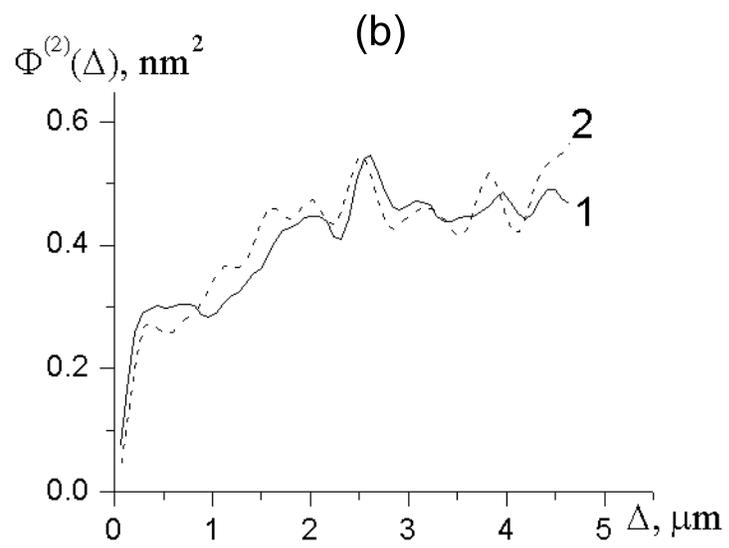

**Figure 12**

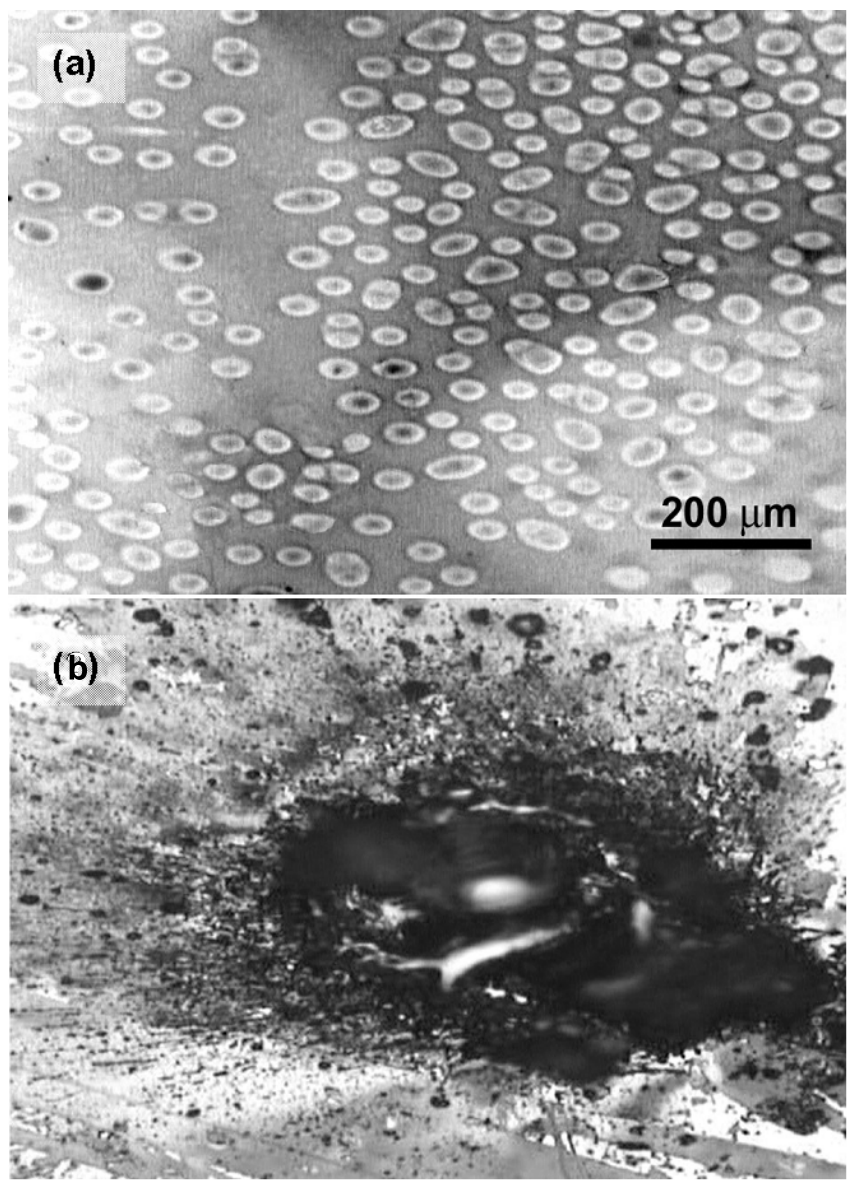

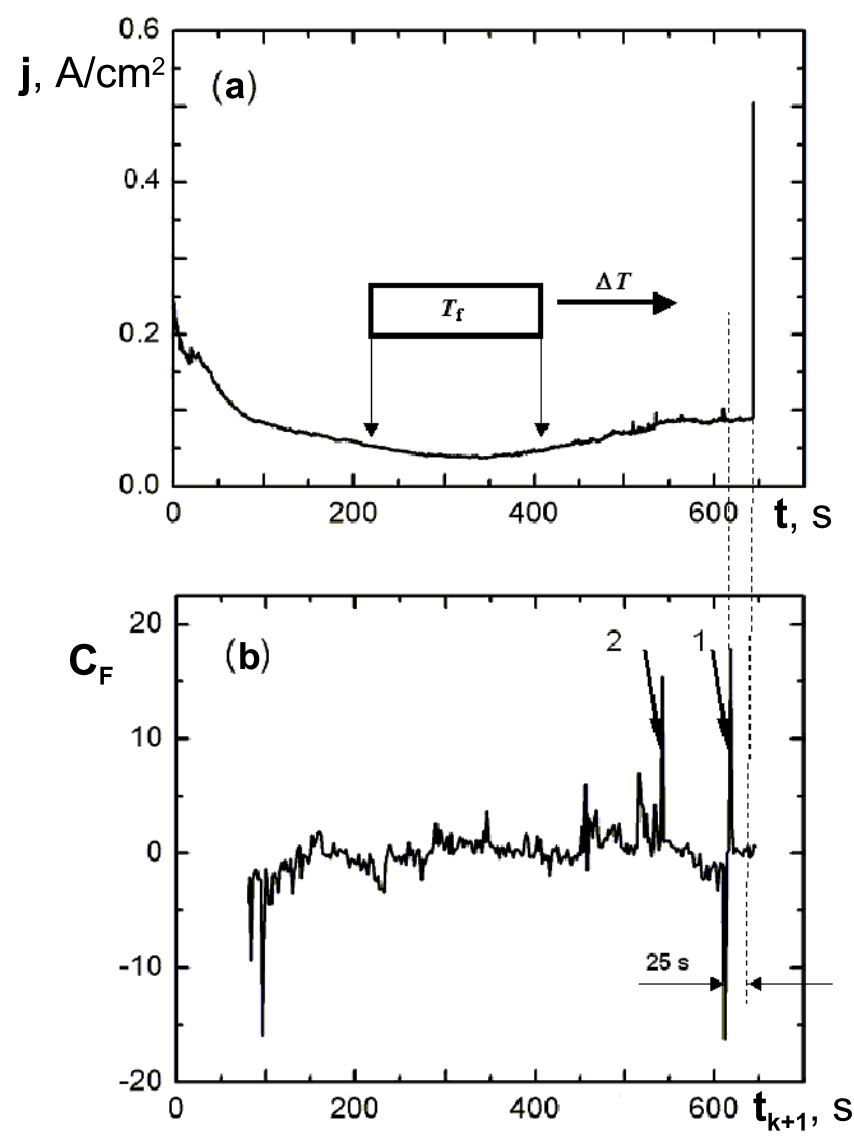

**Figure 14**

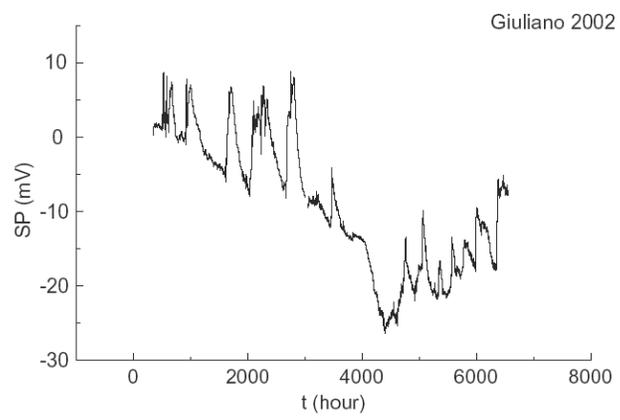

**Figure 15**

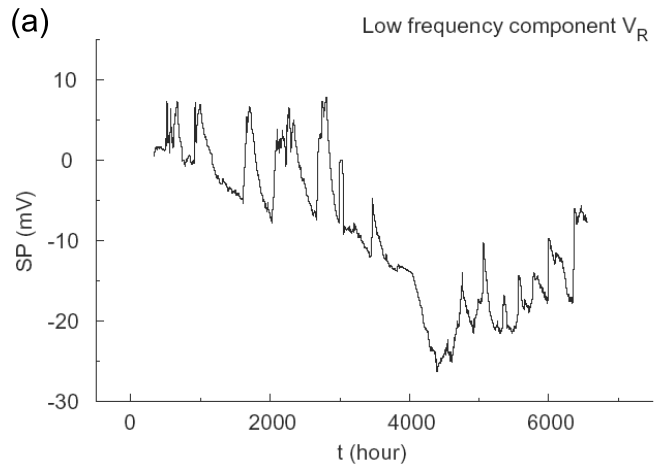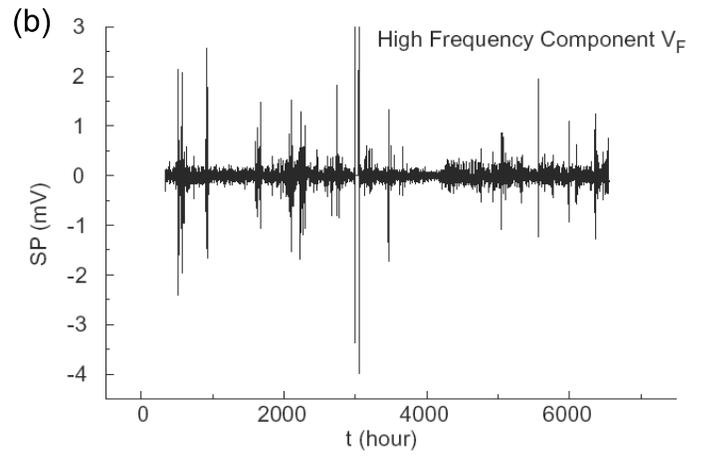

**Figure 16**

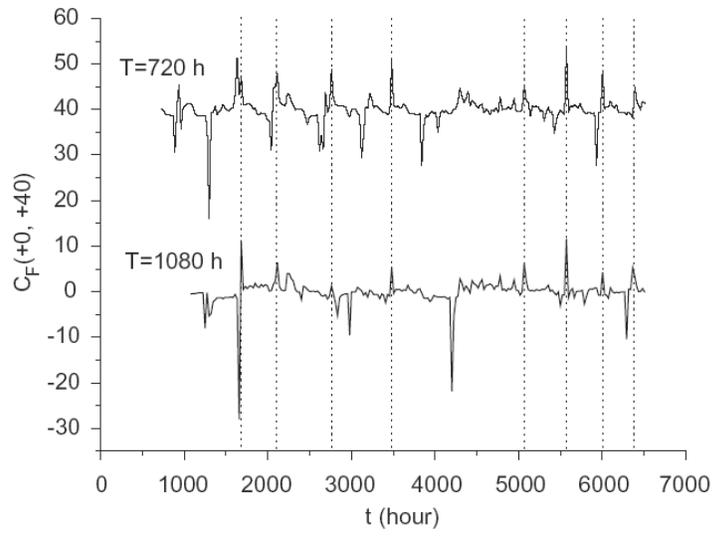

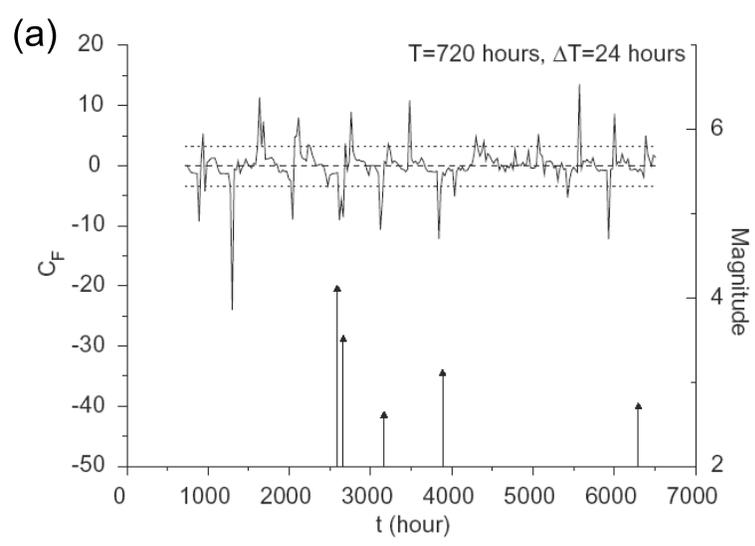
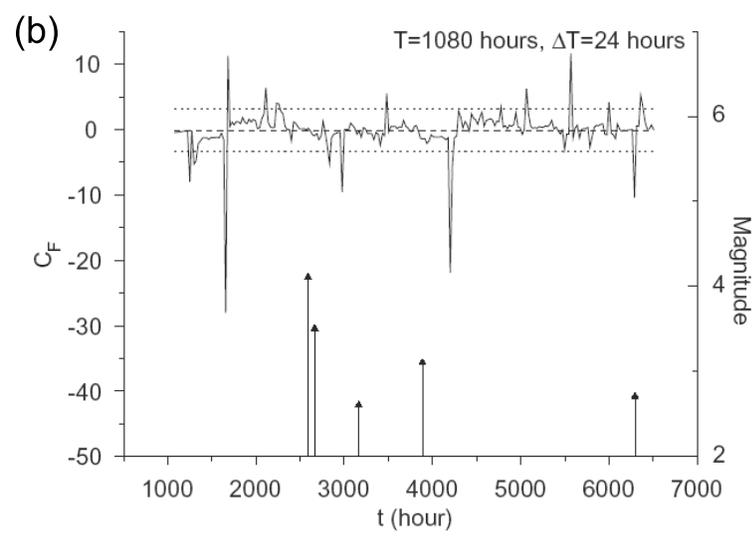

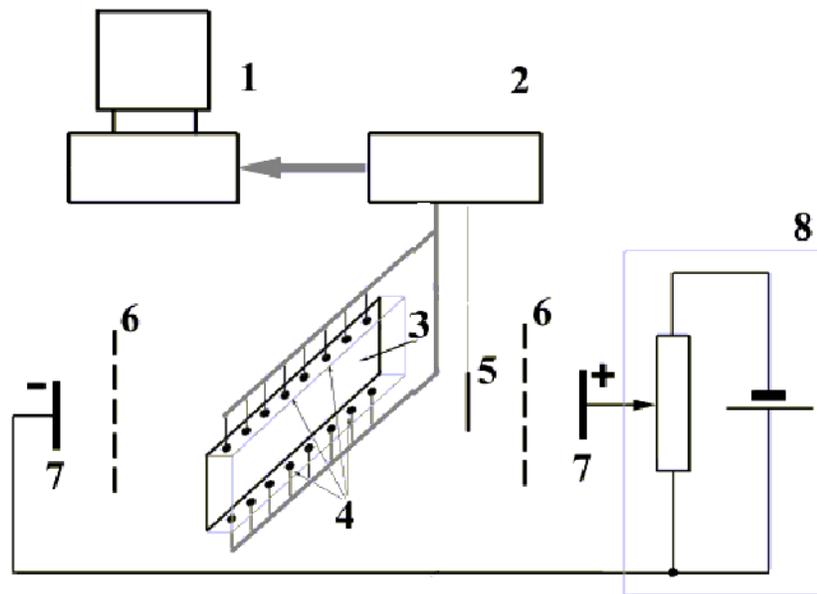

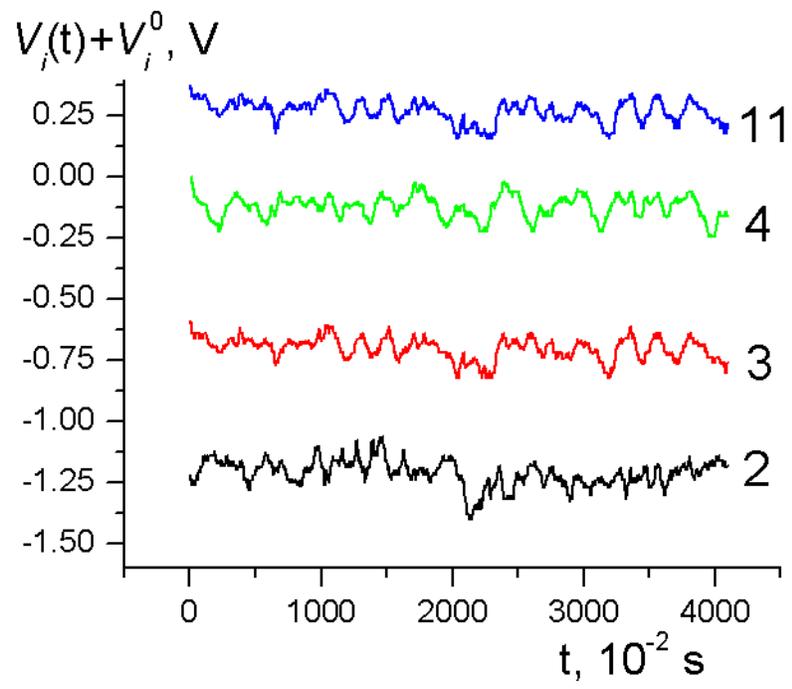

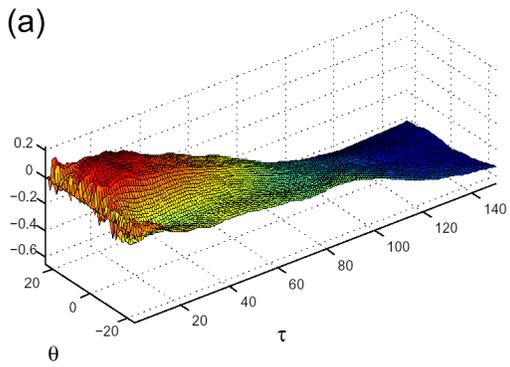
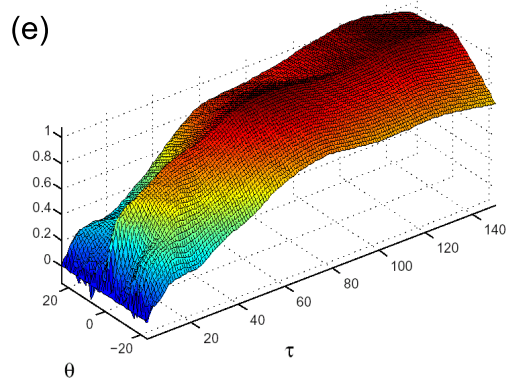
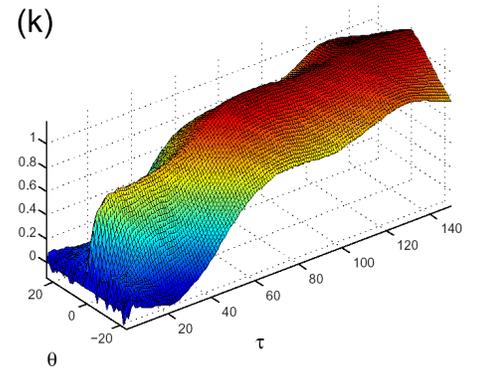
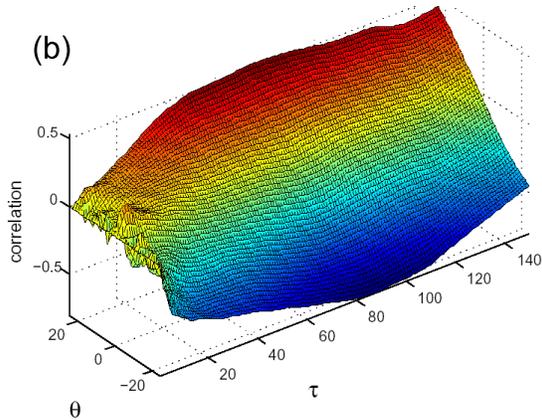
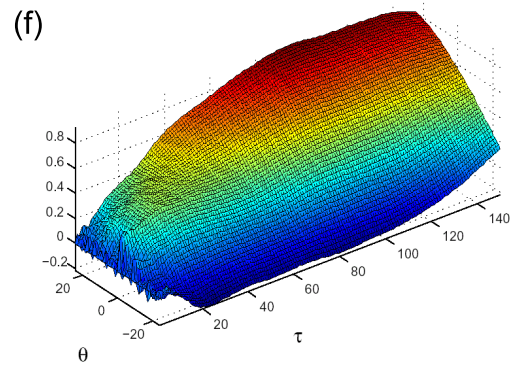
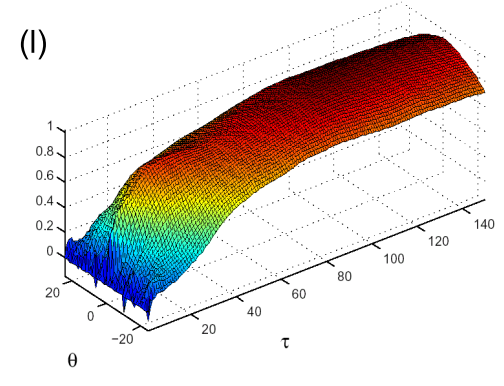
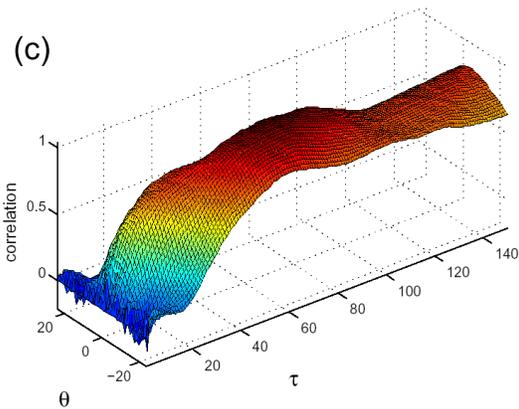
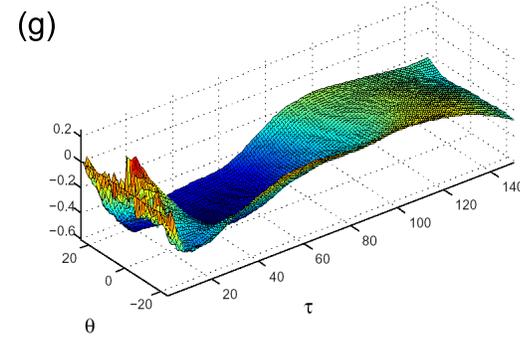
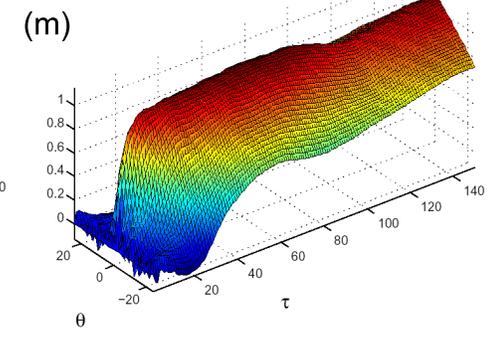
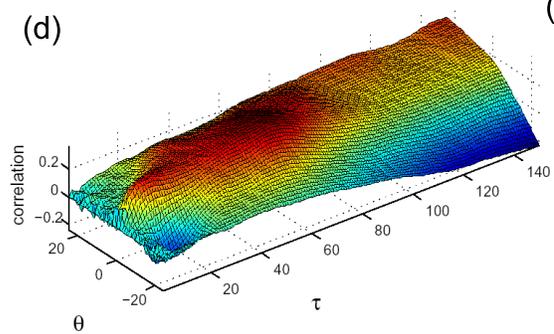
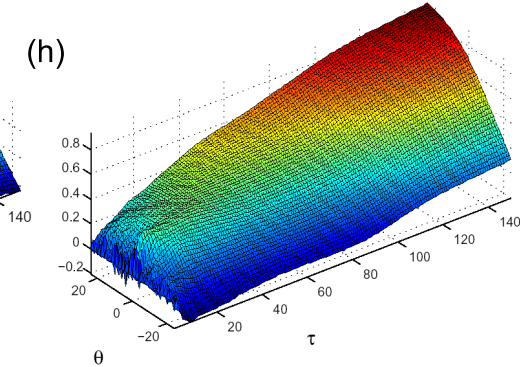
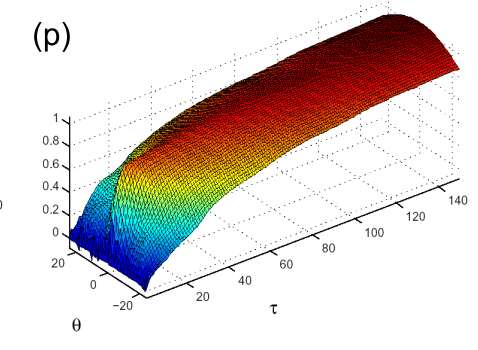

**Figure 21**

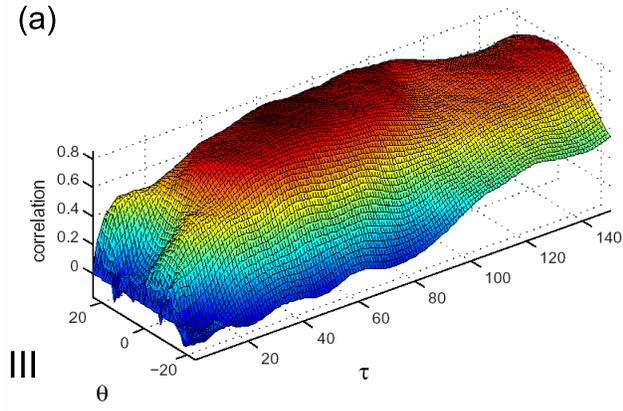
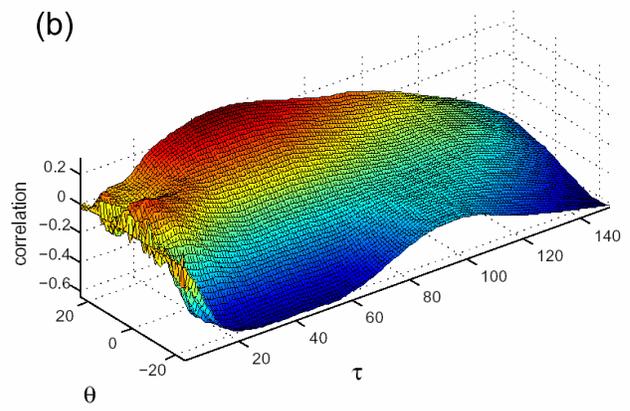